%% file: mastersthesis.tex
\documentclass[12pt,titlepage,a4paper,draft]{report}

\usepackage{amsmath}
\usepackage{amssymb}
\usepackage{epsfig}
\usepackage{latexsym}
\usepackage{graphicx}
\usepackage{epic}
\usepackage{eepic}
\usepackage{setspace}
\usepackage{supertabular}
\usepackage{psfrag}
\usepackage{rotating}
\usepackage{wrapfig}
\usepackage{fancyhdr}
\usepackage[dvips]{color}
\usepackage{cite}
\usepackage{makeidx}
\usepackage[english]{babel}

\psfull

\renewcommand{\chaptermark}[1]%
		{\markboth{#1}{}}
\renewcommand{\sectionmark}[1]%
		{\markright{\thesection\ #1}}
\makeindex

\begin{document}
\addtocontents{toc}{\protect\vspace{4ex}}
\pagenumbering{roman}

\include{newcoms}
\include{cover}
\include{abstracteng}
\include{abstractfin}

\addtolength{\textheight}{-5cm}	
\addtolength{\topmargin}{-2cm}	
\addcontentsline{toc}{chapter}{Acknowledgements}

\include{acknowledgement}

\addtolength{\topmargin}{+2cm}	
\addcontentsline{toc}{chapter}{Table of Contents}
\tableofcontents

\setlength{\baselineskip}{3.5ex}	
\pagestyle{fancyplain}

\include{intro}
\include{clqinfo}
\include{qkd}
\include{analysis}
\include{conclusions}

\setlength{\baselineskip}{1ex}

\addcontentsline{toc}{chapter}{Bibliography}
\bibliographystyle{prsty}
 {\footnotesize{\bibliography{BEC,MyBec}}}
\include{bibliography}
\end{document}

%% file: newcoms.tex
\newcommand{\beq}{\begin{equation}}
\newcommand{\eeq}{\end{equation}}

\newcommand{\beqa}{\begin{eqnarray}}
\newcommand{\eeqa}{\end{eqnarray}}

\newcommand{\bitem}{\begin{itemize}}
\newcommand{\eitem}{\end{itemize}}

\newcommand{\tabref}[1]{Tab.~\ref{#1}}
\newcommand{\Eqref}[1]{Eq.~(\ref{#1})}
\renewcommand{\eqref}[1]{(\ref{#1})}
\newcommand{\figref}[1]{Fig.~\ref{#1}}

\newcommand{\eg}{{\it e.g.}}
\newcommand{\ie}{{\it i.e.}}
\newcommand{\etal}{{\it et al.\/ }}

\newcommand{\mrm}{\mathrm}
\newcommand{\mbf}{\mathbf}
\newcommand{\doo}{\partial}
\newcommand{\bfr}{\mathbf{r}}
\newcommand{\bfp}{\mathbf{p}}
\newcommand{\bfv}{\mathbf{v}}

\newcommand{\Psihat}{\hat\Psi}
\newcommand{\psihat}{\hat\psi}
\newcommand{\Psihatdag}{\hat\Psi^\dag}
\newcommand{\psihatdag}{\hat\psi^\dag}
\newcommand{\Psihatr}{\hat\Psi(\textbf{r})}
\newcommand{\psihatr}{\hat\psi(\textbf{r})}
\newcommand{\Psihatdagr}{\hat\Psi^\dag(\textbf{r})}
\newcommand{\psihatdagr}{\hat\psi^\dag(\textbf{r})}
\newcommand{\firc}{\phi^*(\textbf{r})}
\newcommand{\fir}{\phi(\textbf{r})}
\newcommand{\psir}{\psi(\textbf{r})}

\newcommand{\Psihatdagrp}{\hat\Psi^\dag(\textbf{r}')}
\newcommand{\Psihatrp}{\hat\Psi(\textbf{r}')}

\newcommand{\bfrp}{\textbf{r}'}

\newcommand{\defas}{\mathrel{\raise.095ex\hbox{$:$}\mkern-4.2mu=}}

\newcommand{\nn}{\nonumber}
\newcommand{\half}{\frac{1}{2}}
\newcommand{\pfrac}[2]{\left( \frac{#1}{#2}\right)}
\newcommand{\lrp}[1]{\left( #1 \right)}
\newcommand{\lrpx}[1]{\left[ #1 \right]}
\newcommand{\ti}[1]{\tilde{#1}}

\newcommand{\viite}[4]{\bibitem{#1} #2, \emph{#3,} #4.}
\newcommand{\viitex}[3]{\bibitem{#1} #2, #3.}
\newcommand{\viitek}[4]{\bibitem{#1} #2, \emph{#3} #4.}

\renewcommand{\epsilon}{\varepsilon}
\newcommand{\comment}[1]{}

%% file: cover.tex
\setlength{\topmargin}{-1cm}
\begin{titlepage}
\vspace*{-30pt}
\mbox{}
\hrule
\vspace*{10pt}
\mbox{}\\
\begin{minipage}[t]{15mm} 
\vspace{-14pt}

\epsfig{file=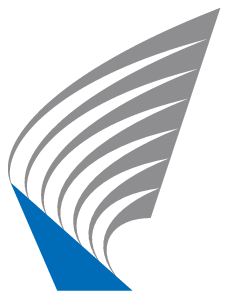, width=15mm}
\end{minipage}
\begin{minipage}[t]{200mm}
%\vspace{4mm}
\begin{tabbing}
Sopiva väli \=TEKNILLINEN KORKEAKOLU12345678901234567 \= 12345667890123456678 \= \kill

\>Helsinki University of Technology \\
\>Department of Engineering Physics and Mathematics \\
\>Laboratory of Physics \\
%\> \today\\
\\

\end{tabbing}
\end{minipage}
\hrule
\begin{tabbing}
fdsa \= blaa blaa\= \kill
\\
\end{tabbing}

\begin{center}
\Large \bf Master's Thesis\\
\end{center}
\begin{tabbing}
\\
\end{tabbing}
\begin{center}
\LARGE \bf Quantum Cryptography Protocol Based on Sending Entangled Qubit Pairs \\
\end{center}
\begin{tabbing}
TEKNILLINEN KORKEAKOLU1234567890123456 \= ERIKOISTYO12\= \kill\\
\\
\\
\end{tabbing}
\begin{center}
\Large \bf Olli Ahonen \\
\end{center}
\begin{tabbing}
\\
\end{tabbing}
\begin{center}
\begin{tabular}{ll}
\large Supervisor:& \large Prof.\ Risto Nieminen\\
\parbox{0mm} {\vspace{2mm} } % \parbox{10mm} {\vspace{-1mm} } \\ 
\large Instructor:& \large Dr.\ Mikko M\"ott\"onen \\
\end{tabular}
\end{center}
\begin{tabbing}
\\
\end{tabbing}
\vspace*{15pt}
\hrule
\large Otaniemi\\
\large March 1st 2007
\end{titlepage}

%% file: abstracteng.tex
\thispagestyle{empty}
{\textsc{Helsinki University of Technology\hfill{Abstract}}}
\vspace{1mm}

\fbox{\fbox{
\begin{minipage}{0.93\linewidth}

\begin{tabular}{p{3cm}l}
\bf{Author:      }  		& Olli Ahonen  \\
\end{tabular}
{\vspace{-5mm}\center{\rule{0.99\linewidth}{1pt}}} \\
\begin{tabular}{p{3cm}l}
\bf{Department:   }  	& Engineering Physics and Mathematics \\
\bf{Major:        }	  	& Materials Physics \\
\bf{Minor:        }  	  	& Software Systems 
\end{tabular}
{\vspace{-3mm}\center{\rule{0.99\linewidth}{1pt}}} \\
\begin{tabular}{p{3cm}l}
\bf{Title:}		  & Quantum Cryptography Protocol Based on Sending\\& Entangled Qubit Pairs \\
\bf{Otsikko:}       & Kietoutuneisiin qubittipareihin perustuva kvantti-\\& salausmenetelm\"a \\ 
\end{tabular}
{\vspace{-3mm}\center{\rule{0.99\linewidth}{1pt}}} \\
\begin{tabular}{p{3cm}l}
\bf{Chair:}	  & Tfy-44 (Materials Physics) \\
\bf{Supervisor:   }	  & Professor Risto Nieminen\\
\bf{Instructor:   }  	  & Dr.\ Tech.\ Mikko M\"ott\"onen\\
\end{tabular}
{\vspace{-3mm}\center{\rule{0.99\linewidth}{1pt}}} \\
\begin{tabular}{p{13.5cm}}
The quantum key distribution protocol BB84, published by C.\ H.\ Bennett and G.\ Brassard in 1984, describes how two spatially separated parties can generate a random bit string fully known only to them by transmission of single-qubit quantum states. Any attempt to eavesdrop on the protocol introduces disturbance which can be detected by the legitimate parties.
\vspace{10pt}

In this Master's Thesis a novel modification to the BB84 protocol is analyzed. Instead of sending single particles one-by-one as in BB84, they are grouped and a non-local transformation is applied to each group before transmission. Each particle is sent to the intended receiver, always delaying the transmission until the receiver has acknowledged the previous particle on an authenticated classical channel, restricting eavesdropping to accessing the quantum transmission one particle at a time. Hence, an eavesdropper cannot undo the non-local transformation perfectly. Even if perfect cloning of quantum states was possible the state of the group could not be cloned.
\vspace{10pt}

We calculate the maximal information on the established key provided by an intercept-resend attack and the induced disturbance for different transformations. We observe that it is possible to significantly reduce the eavesdropper's maximal information on the key---to one eighth of that in BB84 for a fixed, reasonable amount of disturbance.  We also show that the individual access to the particles poses a fundamental restriction to the eavesdropper, and discuss a novel attack type against the proposed protocol.
\end{tabular}
{\vspace{-3mm}\center{\rule{0.99\linewidth}{1pt}}} \\
\begin{tabular}{p{3cm}l}
\bf{Keywords:}   & Quantum cryptography, Quantum key distribution, \\& BB84, Entanglement \\
\end{tabular}
\begin{tabular}{p{0.5\linewidth}l}
{\bf{Pages:}}  	 \hspace{17.5mm} 	  75 & {\bf{Date:}} March 1$^{st}$, 2007\\
\end{tabular}
{\vspace{-5mm}\center{\rule{0.99\linewidth}{1pt}}} \\
\begin{tabular}{p{0.5\linewidth}l}
\bf{Approved:    }  	  & \bf{Location:    }

\end{tabular}

\end{minipage}}}

%entanglement = kietoutuminen tai lomittuminen

%% file: abstractfin.tex
\thispagestyle{empty}
{\textsc{Teknillinen korkeakoulu\hfill{Tiivistelm\"a}}}
\vspace{1mm}

\fbox{\fbox{
\begin{minipage}{0.93\linewidth}

\begin{tabular}{p{3cm}l}
\bf{Tekij\"a:      }  		& Olli Ahonen  \\
\end{tabular}
{\vspace{-5mm}\center{\rule{0.99\linewidth}{1pt}}} \\
\begin{tabular}{p{3cm}l}
\bf{Osasto:   }  	& Teknillisen fysiikan ja matematiikan osasto \\
\bf{P\"a\"aaine:        }	  	& Materiaalifysiikka \\
\bf{Sivuaine:        }  	  	& Ohjelmistoj\"arjestelm\"at 
\end{tabular}
{\vspace{-3mm}\center{\rule{0.99\linewidth}{1pt}}} \\
\begin{tabular}{p{3cm}l}
\bf{Otsikko:}       & Kietoutuneisiin qubittipareihin perustuva kvantti-\\& salausmenetelm\"a \\ 
\bf{Title:}		  & Quantum Cryptography Protocol Based on Sending\\& Entangled Qubit Pairs \\
\end{tabular}
{\vspace{-3mm}\center{\rule{0.99\linewidth}{1pt}}} \\
\begin{tabular}{p{3cm}l}
\bf{Professuuri:}	  & Tfy-44 (Materiaalifysiikka) \\
\bf{Valvoja:   }	  & Professori Risto Nieminen \\
\bf{Ohjaaja:   }  	  & TkT Mikko M\"ott\"onen\\
\end{tabular}
{\vspace{-3mm}\center{\rule{0.99\linewidth}{1pt}}} \\
\begin{tabular}{p{13.5cm}}
C.\ H.\ Bennett ja G.\ Brassard julkaisivat vuonna 1984 BB84:ksi kut\-su\-tun menetelm\"an, jolla toisistaan et\"a\"all\"a olevat osapuolet voivat lu\-o\-da sa\-tun\-nai\-sen bittijonon, jonka vain he tuntevat kokonaan. Me\-ne\-tel\-m\"a hy\"odynt\"a\"a yksiqubittitilojen ominaisuuksia. Kaikki mahdolliset salakuunteluyritykset aiheuttavat h\"airi\"ot\"a, jonka perusteella sala\-kuun\-te\-lu voidaan havaita.
\vspace{10pt}

T\"ass\"a diplomity\"oss\"a tutkitaan uudenlaista BB84:\"a\"an perustuvaa menetelm\"a\"a. Sen sijaan, ett\"a yksitt\"aisi\"a hiukkasia l\"ahetett\"aisiin erikseen kuten BB84:ss\"a, ne ryhmitell\"a\"an, ja kunkin ryhm\"an hiukkasten tilat kiedotaan erityisell\"a muunnoksella. Kietoutuneet hiukkaset l\"ahetet\"a\"an toiselle osapuolelle siten, ett\"a kunkin hiukkasen l\"ahetyst\"a lyk\"at\"a\"an kun\-nes edel\-li\-ses\-t\"a l\"ahetyksest\"a on saatu kuittaus. T\"am\"a rajoittaa salakuuntelun yhteen hiukkaseen kerrallaan, joten salakuuntelija ei pysty perumaan tehty\"a muunnosta t\"aydellisesti. Vaikka t\"aydellinen kvanttitilojen kopioiminen olisi mah\-dol\-lis\-ta, kiedotun ryhm\"an tilaa ei voida kopioida.

\vspace{10pt}

Laskemme eri muunnoksille sieppaus-uudelleenl\"ahetys -hy\"ok\-k\"a\-yk\-sel\-l\"a saatavan maksimaalisen tiedon avaimesta ja kvanttikanavassa aiheutetun h\"airi\"on. Havaitsemme, ett\"a maksimaalista tietoa on mahdollista ra\-joit\-taa huomattavasti, jopa kahdeksasosaan BB84:n vastaavasta arvosta. N\"ayt\"amme, ett\"a salakuuntelun rajoittaminen yhteen hiukkaseen kerrallaan luo perustavanlaatuisen esteen salakuuntelulle. Esit\"amme my\"os uudenlaisen hy\"okk\"ayksen ehdotettua menetelm\"a\"a vastaan.
\end{tabular}
{\vspace{-3mm}\center{\rule{0.99\linewidth}{1pt}}} \\
\begin{tabular}{p{3cm}l}
\bf{Avainsanat:}   & Kvanttikryptografia, Avaimenjako, BB84, \\& Kietoutuminen \\
\end{tabular}
\begin{tabular}{p{0.5\linewidth}l}
{\bf{Sivuja:}}  	 \hspace{17.0mm} 	75 & {\bf{P\"aiv\"ays:}} 1. maaliskuuta 2007\\
\end{tabular}
{\vspace{-5mm}\center{\rule{0.99\linewidth}{1pt}}} \\
\begin{tabular}{p{0.5\linewidth}l}
\bf{Hyv\"aksytty:    }  	  & \bf{Sijainti:    }

\end{tabular}

\end{minipage}}}

%% file: acknowledgement.tex
\chapter*{Acknowledgements}

I would like to kindly thank my instructor Dr.\ Tech.\ Mikko M\"ott\"onen for all his constructive advice and readily available guidance on this Thesis. I am grateful to all the members of the Quantum many-body physics group for creating a positive and pleasant working atmosphere, and to Prof.\ Risto Nieminen and Docent Ari Harju for the marvellous opportunity to work in the group. I would also like to thank the members of the positron group of our laboratory for their invaluable help in the beginning of my academic career.

Special thanks belong to my parents, without whose continual support I cannot imagine having achieved so much in my life, and to my big sister and my little brother for all the joyous moments they have provided during the years. Finally, I wish to thank my fianc\'ee Helena for her unconditional support and the happiness she brings to my life.
%... Many thanks ... Special thanks is obligated to ... I would also like to thank ... Finally, I thank...
\\
\\
\\
\\\emph{\hspace*{7.5cm}``Work like you'd live forever. \\\hspace*{7.5cm} Love like you'd die tomorrow.'' \\\hspace*{8cm}--Seneca}
%\\\hspace*{7cm}Yep''}
\\
\\
\\
\\
\\
\\
\\
\\
\\
\\
\\
\\
\\
\emph{Otaniemi, March 1st 2007}
\\
\\
\\
Olli Ahonen

%% file: intro.tex
\chapter{Introduction}\label{ch:intro}

\pagenumbering{arabic}
In human societies, the desire of two parties to communicate in secret dates back at least as far as the first known societies themselves \cite{dkahn}. When the two parties are not in perfect isolation, that is, when the messages they exchange may become available to outsiders, the best known technique to achieve secrecy is \emph{cryptography}. Cryptography is about concealing the meaning of communicated messages from any unintended recipients. This is traditionally achieved by the following scheme: The legitimate parties share a relatively small amount of secret information, a \emph{key}, based upon which the sender chooses a transformation, and the receiver another transformation that perfectly undoes the effect of the former. Each message is transformed, i.e. \emph{encrypted}, by the sender before sending the message, and then re-transformed, i.e. \emph{decrypted}, by the receiver to recover the original message. Only receivers in possession of the correct key know exactly which decrypting transformation to apply.

Today, secrecy of communication is not compromised by lack of secure encryption-decryption schemes, but is rather hindered by the complicated problem of delivering the needed encryption-decryption keys safely. This fact is nicely exemplified by a cryptographic protocol known as the one-time pad, first proposed by J.\ Mauborgne \cite{stallings}. The one-time pad scheme requires that the two communicating parties share in advance a secret key that is as long as the message they wish to transmit. Assuming that the key and the message are in binary form, i.e., strings of zeroes and ones, both the sender and the receiver apply the bitwise exclusive-or (XOR) operation: Each bit in the message to be sent is flipped if and only if the corresponding bit in the key has value one. After reception, the receiver performs exactly the same operation, and recovers the original message. As long as the key is never reused, the one-time pad is unbreakable. That is, without knowledge of the key, the communication is perfectly secret.

The one-time pad keys are far from a relatively small amount of secret information, and hence agreeing on them is in general a daunting task. Therefore, the one-time pad as such is not of much practical value, despite its extreme security. Instead of insisting on perfect security, contemporary protocols use a rather short key many times but involve transformations more complicated than a single bitwise XOR. These protocols are, in principle, vulnerable to careful analysis of a large amount of captured, although encrypted, communication. With key lengths of 128 to 256 bits, however, cryptographic protocols employing, e.g., the Advanced encryption standard (AES) algorithm, are considered secure enough \cite{cnss}.

One way of solving the problem of key distribution is to utilize asymmetric \emph{public-key encryption} which is in widespread use today. In public-key encryption schemes, two parties desiring secret communication need not share any secret information in advance. The intended receiver of the messages can give the sender a \emph{public key} which the sender uses to encrypt messages. Messages encrypted with the public key can only be decrypted with a corresponding \emph{private key}. Public-key encryption is usually used to exchange keys for symmetric encryption-decryption schemes that use only one key, for example, AES-based protocols.

In public-key encryption schemes, deducing the private key from the public key is generally considered \emph{hard}, i.e., the deduction would take an overwhelming amount of time on any computer. However, this conjecture has never been proved. Thus, it is possible that this deduction can be performed in a reasonable amount of time. Moreover, it is known that the public-key to private-key deduction is feasible on a large-scale \emph{quantum computer}, but the demanding task of constructing such a device remains yet to be accomplished. Because of the possibility of recording encrypted transmissions, the security of not only future but also all past public-key-encrypted communication is based on a conjecture of the difficulty of the key deduction.

The history of \emph{quantum cryptography} can be considered to begin in 1969. By then, the peculiar properties of quantum physics, discovered in the early decades of the 20th century, had not only been developed into a rigorous theory, but also increasingly adopted by scientists. In 1969, Stephen Wiesner introduced\footnote{Wiesner's proposition was not, however, published until 1983 \cite{wiesner}.} the idea of forgery-proof quantum banknotes \cite{galindo}. These banknotes would contain a serial number encoded by the issuing bank into quantum states of individual particles. By the laws of quantum mechanics, anyone unaware of the details of the encoding could not produce copies of the banknotes. Although this idea of uncloneable sequence of numbers is not an essential primitive or function in quantum cryptography, the next milestone in the field owes much to its insight \cite{afterdinner}.

This milestone is the \emph{quantum key distribution protocol} of C.\ H.\ Bennett and G.\ Brassard published in 1984 \cite{bb84}. The protocol, referred to as BB84, describes how two spatially separated parties can generate a random bit string, a sequence of zeroes and ones, fully known only to them. The protocol exploits properties of single quanta predicted by quantum physics. The sender transmits a sequence of individual particles, e.g., photons, each in one of four equally probable quantum states. Any eavesdropping on the states of the particles between the sender and the intended receiver inevitably introduces disturbance to the transmission, which can later be detected by the legitimate parties. In addition, an eavesdropper has only bounded information on the established bit sequence. These properties are guaranteed by the laws of physics: An eavesdropper can only gain knowledge on the transmission by measurements. The states of the particles are chosen so that any physically conceivable measurement will never provide the eavesdropper with full knowledge on the transmission, and will almost certainly disrupt the transmission. The protocol requires that the sender and the receiver can communicate via an authenticated\footnote{On an authenticated channel an outsider cannot pose as a legitimate user.} classical channel, e.g., a phone line. The classical channel can be assumed public, e.g., wire-tapping is allowed on a phone line.

Since it was first introduced, several variations of and modifications to the BB84 protocol have been proposed. These include, for example, modifying the number of states allowed to the transmitted particles \cite{mod2state,mod6state}, adjusting the probabilities of the individual states of the particles \cite{modbasesprob}, transmitting the individual states on more than one particle \cite{modsothers11}, and exploiting quantum-mechanical spatial superposition \cite{modsothers24}.

The so-called Einstein-Podolsky-Rosen (EPR) protocols offer a different, yet in many ways equivalent, approach to quantum key distribution \cite{modEPR}. In EPR protocols, the individual particles are not transmitted from sender to receiver but emanate from a separate source. This source always emits two particles in an \emph{entangled} state, one particle for each party of the protocol. Entangled particles exhibit correlations independent of their spatial distance. These correlations are exploited by the participants of the protocol to establish a secret random bit sequence. Once again, it is possible to detect if anyone has tampered with the source or the particles before their reception. An authenticated classical channel is needed in this protocol, as well.

As such, none of the quantum-cryptographic protocols mentioned above provide the participants with an error-free perfectly secret bit sequence---one that no-one else has any knowledge of. They rather allow the two parties to share a bit sequence, and guarantee that no-one else has information on the sequence above some fixed value. Furthermore, their sequences do not match perfectly because of unavoidable noise and possible eavesdropping on the transmitted quantum states. A decent amount of errors can be corrected safely by communication over a public classical channel. Moreover, the participants may perform \emph{privacy amplification}: they exchange further information over the classical channel while shortening their bit sequence, and thus reduce any eavesdropper's knowledge of the sequence to an arbitrarily low value. Hence, the legitimate parties can finally obtain a shared, perfectly secret, error-free bit sequence.

Above, we discussed how quantum cryptography can be used to send secret random bit strings between two parties. What is the value of this capability in terms of secret communication? Indeed, if they are strictly random, these secret shared bit sequences provided by the quantum protocols described above can---after error correction and privacy amplification---be directly used as keys in classical symmetric cryptography already in everyday use. This essential and non-trivial key-distribution function is what quantum cryptography in its modern form is most suited for. Hence, quantum cryptography and quantum key distribution are often used synonymously. The practical applications of these protocols can safely replace risky public-key-encrypted key distribution methods.

Quantum cryptography became reality in 1992, as C.\ H.\ Bennett \emph{et al.}~experimentally implemented the BB84 protocol for the first time \cite{bb84implemented}. The protocol was completed between parties 30 cm apart---a distance not of much interest in a commercial application. Since the first experimental realization, quantum key distribution has been succesfully carried out over distances of tens of kilometers \cite{freespace13km, qber034}. However, quantum cryptography is still plagued by the difficulty of realizing the protocol over longer distances. Practical applications invariably use photons as the carrier of the quantum states. Photons may be transported either in optical fibre or in free space, i.e., earth's atmosphere. For either choice, the rapid decay of faint light pulses prohibits secure quantum cryptography over distances above 100 km. \cite{bouwmeester, gisin}

In this Master's Thesis, a novel modification to the BB84 protocol is analyzed. This modification is outlined as follows: Instead of sending the single particles one-by-one, they are grouped and a transformation coupling the particles is applied to each group before transmission. This transformation is assumed to be known by the sender and receiver, as well as any eavesdropper. After the transformation, the particles are sent to the intended receiver, but always delaying the transmission until the receiver has acknowledged receiving the previous particle on the authenticated classical channel. This restricts eavesdropping to accessing the quantum transmission one particle at a time. But because a transformation involving a group of particles was applied, the eavesdropper generally cannot undo that transformation perfectly. Moreover, even if the eavesdropper was capable of cloning the states of the individual particles for herself\footnote{In fact, we show in Sec.~\ref{sec:cloning} that this is not possible.}, she could not clone the state of the entire group formed by the sender. The legitimate receiver, on the other hand, can undo the transformation used by the sender because he or she has simultaneous access to all the particles of a group. Not all quantum transformations involving a group of particles are such that they cannot be reversed one particle at a time; the transformation has to be \emph{non-local}. The aim of this Thesis is to find out exactly which non-local transformation allows the legitimate users of the modified protocol to gain the maximal advantage over an eavesdropper. Furthermore, we restrict ourselves to the case where the group size is two particles, i.e., the particles are handled in pairs.

Chapter \ref{ch:clqinfo} reviews the concepts and results of classical and quantum information theory needed in subsequent chapters. Chapter \ref{ch:qkd} defines and describes in detail the aspects of quantum key distribution relevant to our studies. The analysis of the modified protocol is presented in Ch.~\ref{ch:analysis} which includes the obtained results. Finally, Ch.~\ref{ch:conclusions} concludes this Thesis with a summary and suggestions for future research.

%% file: clqinfo.tex
\chapter{Classical and Quantum Information}
\label{ch:clqinfo}

This chapter presents mathematical tools of classical and quantum information needed in our studies. The reader is assumed to be familiar with elementary quantum mechanics which will not be discussed here. In the first section, we briefly review some useful results of probability theory. In Sec.~\ref{sec:entropy}, we define information-theoretic \emph{entropy} and its derivative, \emph{mutual information}. Entropy is indisputably the central concept in classical information theory. Mutual information enables us to give quantitative expression to statements such as: ``An adversary has knowledge on a secret key.'' Section \ref{sec:errcorr} discusses correcting errors in transmitted data due to an imperfect channel between two parties. Error correction is accomplished by exchanging further information about the erroneous data.

The latter part of this chapter discusses topics specific to quantum information. Sections \ref{sec:qubit} and \ref{sec:entanglement} introduce the quantum analog of the bit, namely the \emph{qubit}, and the properties that most notably distuinguish qubits and classical bits. A short discussion concerning the physical realizations of a qubit is also included. Section \ref{sec:measurement} reviews means to read existing qubits, i.e., quantum-mechanical measurements. Finally, copying, or cloning, of quantum information is discussed in Sec.~\ref{sec:cloning}.
%korjaus29.12. entanglement-kappale mainittu JA pilkku poistettu namely jälkeen

\section{Elementary probability theory}
The theory of probability can be formulated on the basis of the notion of a \emph{random variable}. A random variable $X$ may assume a number of values, and the value $X$ has assumed is denoted by $x$. The probability with which $X$ assumes the value $x$, is denoted by $p(X=x)$, which may also be written $p(x)$. If the possible values $X$ may take are $x_1,x_2,...,x_n$, the \emph{probability distribution} of $X$ is $p_X := (p(x_1),p(x_2),...,p(x_n)) := (p_1,p_2,...,p_n)$. Random variables relevant to this Thesis always take their value from a finite set. The probability that two distinct random variables $X$ and $Y$ assume values $x$ and $y$, respectively, is denoted by $p(X=x \mathrm{~AND~} Y=y)$ or, in short, $p(x,y)$. The probability that $X$ assumes $x$ or $Y$ assumes $y$, or both, is denoted by $p(X=x \mathrm{~OR~} Y=y)$. The notation also generalizes to more than two random variables.

For any two random variables $X$ and $Y$, and their respective outcomes $x$ and $y$, we have
\begin{equation}
\label{eq:probor}
p(X=x \mathrm{~OR~} Y=y) = p(x) + p(y) - p(x,y) \; .
\end{equation}
\emph{Conditional probability} is defined by
\begin{equation}
\label{eq:condprob}
p(x|y) := \frac{p(x,y)}{p(y)} \; ,
\end{equation}
and gives the probability that $X$ assumes value $x$, when it is known that $Y$ has taken value $y$. When $p(y)=0$, we define $p(x|y)=0$. When $Y$ has no effect on the outcome of $X$, $p(x|y) = p(x)$, and vice versa. Hence, two random variables are \emph{independent} if and only if $p(x,y)=p(x) p(y)$. Equation (\ref{eq:condprob}) implies \emph{Bayes' theorem} which states that
\begin{equation}
\label{eq:bayes}
p(x|y) = p(y|x) \frac{p(x)}{p(y)} \; .
\end{equation}
This expression is often amended by writing $p(y)$ in the form given by another direct consequence of Eq.~(\ref{eq:condprob}), the \emph{law of total probability}:
\begin{equation}
\label{eq:totalprob}
p(y) = \sum_x p(y|x) p(x) \; .
\end{equation}
From these equations, the following rules can be derived for random variables $A$, $B$, and $C$ and their respective outcomes $a$, $b$, and $c$:
\begin{eqnarray}
\label{eq:probsum}
p(a|b)	& =	& \sum_c p(a,c|b) \; , \\
\label{eq:cond1}
p(a,b|c)	& =	& p(a|b,c) p(b|c) \; , \\
\label{eq:cond4}
p(a|b,c)	& =	& p(b|a,c) \frac{p(a|c)}{p(b|c)} \; .
\end{eqnarray}

\section{Shannon entropy and mutual information}
\label{sec:entropy}
Entropy is described with respect to an \emph{information source}, or equivalently, a random variable. Entropy quantifies the average information gain per use of an information source, or per instance where we let a random variable $X$ assume a value in accordance with its probability distribution $p_X$. Entropy describes our uncertainty of the random variable \emph{before} it has assumed its value, or alternatively, how many \emph{units of information} we have acquired \emph{after} we have learned its value. These two views are equivalent.

The choice for the unit of information is embedded in the definition of entropy. In this Thesis, the unit of classical information is always a bit. Hence, logarithms are always taken to base two: $\log(\,\cdot\,) := \log_2(\,\cdot\,)$, unless otherwise stated.

The Shannon entropy of a random variable $X$ with probability distribution \linebreak$(p_1,p_2,...,p_n)$ is defined by
\begin{equation}
\label{eq:entropy}
H(X) := H(p_1,p_2,...,p_n) = -\sum_{j=1}^n p_j \log p_j \; .
\end{equation}
An impossible event should not contribute to entropy, and we therefore define $0\log(0):=0$. Note that entropy achieves its minimum, zero, if one of the probabilities $p_j = 1$. For a given number of possible outcomes $n$, entropy is maximized if the probability distribution is flat, i.e., $p_j = \frac{1}{n}$ for all $j \in \{1,2,\ldots,n\}$. The maximal value is $\log n$. The entropy of the simplest random variable that is still meaningful, one having only two possible outcomes with probabilities $p$ and $1-p$, is known as \emph{binary entropy}. Its explicit formula is
\begin{equation}
H_{\mathrm{bin}}(p) :=  - p \log p - (1-p) \log (1-p) \;.
\end{equation}

To quantify the average combined information gain of the outcomes $x$ and $y$ of two random variables $X$ and $Y$, we define the \emph{joint entropy} by
\begin{equation}
\label{eq:jointentr}
H(X,Y) := -\sum_{x,y} p(x,y) \log p(x,y) \; .
\end{equation}
The above equation implies that $H(Y,X) = H(X,Y)$.

The mutual information of two random variables $X$ and $Y$ describes how much information they have in common, and is defined by
\begin{equation}
\label{eq:mi}
I(X,Y) := H(X\!:\!Y) := H(X) + H(Y) - H(X,Y) \; .
\end{equation}
It is clear that $I(Y,X) = I(X,Y)$. It also holds that $I(X,Y) = 0$ if and only if $X$ and $Y$ are independent. Equation (\ref{eq:mi}) can be intuitively justified as follows: The first two terms, $H(X)$ and $H(Y)$, represent the information content of $X$ and $Y$. Any information common to them is included in both terms, while their individual information content is included only in the respective term. It follows that common or joint information is counted twice and non-common information once. Therefore, the third term, $-H(X,Y)$, subtracts the individual and common information of the variables, leaving only their mutual information.

\section{Error correction of shared data}
\label{sec:errcorr}
Consider the following scenario: Two parties, Alice and Bob, share data as a result of some completed protocol. For simplicity, we assume the data is a string of bits. The string contains errors, e.g., Alice has 011010 where Bob has 010010, there is an error in the third bit. Alice and Bob wish to correct all errors by exchanging further data, disclosing as little information as possible about the string to any other parties. The theoretical minimum of bits that Alice and Bob have to exchange is
\begin{equation}
\label{eq:errcorrbits}
r = n \big(-p \log p - (1-p)\log (1-p)\big) = n H_{\mathrm{bin}}(p) \; ,
\end{equation}
where $n$ is the length of the string and $p$ is the individual probability of error for each slot in the string \cite{bouwmeester}. The result is based on Shannon's noiseless coding theorem, for which unfortunately only a non-constructive proof is known \cite{welsh}. That is, we know an error-correcting protocol requiring the exchange of only $r$ bits is possible, but we do not know the details of any such protocol.

\subsection{A simple error-correcting protocol}
As an example of error correction, we describe a very simple and inefficient error-correcting protocol. In this context, inefficiency means that parts of the original data must be discarded, and that the protocol exchanges much more than $r$ bits for a realistic value of $p$. The protocol works as follows: Alice randomly chooses pairs of slots in the string, and sends Bob the slot numbers and her XOR value of the bits in the two slots. Bob replies with his XOR value of the corresponding bits. If Alice and Bob's XOR values match, they keep the bit in the first slot and discard the second-slot bit. If their XOR value does not match, they discard both bits. The longer Alice and Bob iterate this procedure, the smaller the probability of an error in their string becomes. \cite{gisin}

\subsection{The Cascade error-correcting protocol}
As a second example of error-correction schemes, we review a protocol significantly more advanced than that described above. The so-called Cascade protocol, presented by G.\ Brassard and L.\ Salvail, is practical and efficient, preserves the length of the original data, and nearly achieves the bound in Eq.~(\ref{eq:errcorrbits}) \cite{optimalerrcorr}. Let $A = (A_1,A_2,\ldots,A_n)$ be Alice's and $B = (B_1,B_2,\ldots,B_n)$ Bob's bit string, where $A_j,B_j \in \{0,1\}$.

In the first \emph{pass} of the protocol, Alice and Bob choose an integer $k_1$, and group their strings into blocks of length $k_1$. Both Alice and Bob compute the parity of each block, i.e., sum up the $k_1$ bits of each block modulo 2. Alice then sends the parities of her blocks to Bob. When Bob discovers that the parity of one of his blocks, $B_l$, differs from Alice's corresponding parity, Bob knows that there is an odd number of bit errors in block $B_l$. Now Alice and Bob initiate the Binary protocol, in which they interactively and recursively find the erroneous bit in the block. The Binary protocol works as follows: Alice sends Bob the parity of the first half of the block. Bob compares this to the parity of the first half of his block. If these parities agree, the error must be in the second half, and if they disagree, the error is in the first half. Binary is then applied to the block half containing the error. The recursion ends and one erroneous bit is corrected, when the block size reaches a small enough value. At this point, the first pass is complete, leaving all of Bob's blocks with either zero or an even number of bit errors.

In each pass $j>1$ Alice and Bob agree on a random shuffling of the bits, and apply the steps of the first pass to the string, with block size $k_j$. Each time an error is corrected, there must be previously treated blocks, whose amount of bit errors just changed from even to odd. Let $\mathcal{K}$ denote the set of these blocks. Alice and Bob apply the Binary protocol to the smallest block in $\mathcal{K}$, and repeat this until all blocks in $\mathcal{K}$ are treated. This ends pass $j$. The block sizes $k_i$ and the number of executed passes must be optimized so that the probability of any remaining errors is small enough and at the same time the leakage of information to outsiders is minimal. The parameters depend on the bit error rate $p$.

\section{The unit of quantum information}
\label{sec:qubit}
Classical information is most conveniently stored and manipulated as bits, and, in classical computer science, the bit is the most commonly used unit of information~\cite{wikibit}. The bit has a quantum analog, the qubit\footnote{To distinguish a classical bit from a qubit, it is sometimes labeled 'cbit'.}, which in turn is the most widely used unit of quantum information. In principle, any quantum system that has at least two states can be considered a qubit: Given any multilevel system, two of its states are simply labeled $|0\rangle$ and $|1\rangle$, in the Dirac bracket notation. This two-state subsystem constitutes the qubit. Furthermore, one can make the distinction between \emph{logical} and \emph{physical qubits}. A logical qubit is a mathematical concept, an aid in the theoretical discussion of quantum information processing, whereas physical qubits actually span the suitable physical subspaces for logical qubits. In this Thesis, the word qubit refers to a logical qubit.

\subsection{Realization of a qubit}
\label{sec:realqubit}
D.\ DiVincenzo has compiled a list of criteria for a particular physical system in order for it to be useful as a qubit in a functioning quantum computer \cite{divincenzo}. This often-quoted list is known as the DiVincenzo criteria, and it applies in many respects to quantum cryptography, as well. The first criterion states that the system should be scalable and the qubits should be well characterized. That is, the relevant physical parameters of the qubit should be known, e.g., the internal Hamiltonian, states of the physical qubit, and couplings external fields. Secondly, it should be possible to initialize a relevant array of qubits to a known, low-entropy state, e.g., $|00...\rangle$. The third criterion concerns aspects of the protocol presented in this Thesis. It states that decoherence of qubit states should progress much slower than the unitary operations, that is, quantum gates\footnote{Unitary operations targeted to qubits are called gates in quantum computation. They are the quantum analogy of logical gates targeted to cbits.}, applied to the qubits. Obviously, computation is not possible if the information contained in the qubits irreversibly leaks into the environment between elementary operations. The fourth criterion is related to the theory of quantum computation. It states that for the proposed realization of the qubit there should exist a set of quantum gates that is universal, i.e., a set with which any unitary operation is achievable. Finally, the result of a quantum information processing task should be attainable: The fifth criterion of DiVincenzo requires the individual qubits to be measurable with high fidelity.
%korjaus 5.1.07 gates-selitys

In addition to the list presented above, DiVincenzo points out two requirements for successful quantum communication. A realization of the qubit, when used in communication, must be such that it is possible to convert stationary qubits, used for local computation or data storage, into flying qubits that are the ones exchanged by two communicating parties. The conversion from flying to stationary qubits should also be achievable. There is, of course, no restriction on the flying and stationary qubits being the same physical system. In addition, it should be possible to transmit flying qubits between specified locations with a low error rate.

For instance, the polarization of a photon can be treated as a flying qubit. The vertical polarization, relative to a fixed frame of reference, may be chosen to be the state $|0\rangle$, and the horizontal polarization to be the state $|1\rangle$. For further example, Elliott \textit{et al.} \cite{arxivelliott} present a network, in which secure quantum key distribution is achieved with photons carrying the flying qubits. The authors report an error rate of 6-8\% for the qubits.
%tähän voisi laittaa lisää viitteitä miten hyvin kriteerit 6 ja 7 täyttyy fotoneilla

\subsection{Superposition}
What really distinguishes quantum bits from classical bits, is the possibility of \emph{superposition}: The qubit may occupy state $|0\rangle$, state $|1\rangle$, or a linear combination of these two, which is always of the form
\begin{equation}
\label{eq:superpos}
|Q\rangle = \alpha |0\rangle + \beta |1\rangle \; ,
\end{equation}
where the complex numbers $\alpha$ and $\beta$ satisfy $|\alpha|^2+|\beta|^2 = 1$ to ensure that the state is normalized. For $\alpha,\beta \neq 0$, the state of the qubit is said to be a superposition of $|0\rangle$ and $|1\rangle$.

\section{Entanglement}
\label{sec:entanglement}
Not only single qubits can occupy a superposition state, but any number of qubits can jointly occupy a superposition state. If this many-qubit superposition state is inexpressible as a tensor product of individual qubit states, the qubits are said to occupy an \emph{entangled} state. Entanglement has no classical analogy, since collections of classical bits in any state are always expressible by specifying the state of its components individually, whereas this is not true for collections of quantum bits. For example, the two-qubit normalized state\footnote{Notation: $|\Psi_1\rangle \otimes |\Psi_2\rangle := |\Psi_1\rangle |\Psi_2\rangle := |\Psi_1 \Psi_2\rangle$, where $\otimes$ is the Kronecker product.} $|\Psi\rangle = \frac{1}{\sqrt{2}}(|00\rangle + |01\rangle)$ is not an entangled state, because it can be decomposed as $|\Psi\rangle = \frac{1}{\sqrt{2}} |0\rangle\otimes(|0\rangle + |1\rangle)$. In contrast, the state $|\Phi\rangle = \frac{1}{\sqrt{2}}(|00\rangle + |11\rangle)$ is entangled, as it cannot be written in a tensor-product form.
%Entanglement plays a crucial role in the protocol analyzed in this Thesis, as will be seen in Ch.~\ref{ch:analysis}.

Entanglement can be considered as an elementary resource in quantum information processing \cite{nielsenchuang, entanglB}, and hence one should be able to express the amount of entanglement in a given quantum state. However, there is no general agreement on how to quantify entanglement. As an example of entanglement measures we briefly review the \emph{entanglement of formation} for a qubit pair, summarizing the work of W. K. Wootters \cite{concurrence}. An extensive list of entanglement measures is given, for example, in Ref.~\cite{entanglB}.

\subsection{Entanglement of formation and concurrence}
Suppose a bi-partite quantum system consisting of subsystems $A$ and $B$ is in state $\rho$. State $\rho$ has a number of pure-state decompositions $\{p_i,|\psi_i\rangle\}$, for which
\begin{equation}
\rho = \sum_i p_i |\psi_i\rangle\langle\psi_i| \; .
\end{equation}
For each pure state $|\psi_i\rangle\langle\psi_i|$ of the system, the entanglement of formation is
\begin{equation}
E(\psi_i) := - \mathrm{Tr}(\rho_A \log \rho_A) =- \mathrm{Tr}(\rho_B \log \rho_B) \; ,
\end{equation}
where $\rho_A = \mathrm{Tr}_B (|\psi_i\rangle\langle\psi_i|)$, i.e., the reduced density operator of subsystem $A$, and similarly for $\rho_B$. Entanglement of formation of $\rho$ is then
\begin{equation}
E(\rho) := \min \sum_i p_i E(\psi_i) \; ,
\end{equation}
where the minimization is taken over all pure-state decompositions of $\rho$. To derive an explicit analytic formula for $E$, Wootters defines \emph{concurrence}, another measure of entanglement, as
\begin{equation}
C(\psi) := |\langle\psi| \widetilde{\sigma}_y |\psi^*\rangle| \; ,
\end{equation}
where $\widetilde{\sigma}_y$ is the conventional Pauli spin matrix $\sigma_y$, if $|\psi\rangle$ is a single-qubit state, and $\sigma_y^{\otimes n}$, if $|\psi\rangle$ is an $n$-qubit state.

It holds that $E(\psi) = \mathcal{E}[C(\psi)]$, where the function $\mathcal{E}$ is given by
\begin{equation}
\mathcal{E}(C) = H_{\mathrm{bin}} \bigg(\frac{1+\sqrt{1-C^2}}{2}\bigg) \; .
\end{equation}
It is shown in Ref.~\cite{concurrence} that
\begin{equation}
E(\rho) = \mathcal{E}[C(\rho)] \; ,
\end{equation}
where
\begin{equation}
C(\rho) = \max\{0,\lambda_1 - \lambda_2 - \lambda_3 - \lambda_4\} \; ,
\end{equation}
where the $\lambda_i$ are the eigenvalues, in decreasing order, of a hermitian matrix $R := \sqrt{\sqrt{\rho} \tilde{\rho} \sqrt{\rho}}$, where tilde denotes the operation of complex conjugation followed by the operation of $\widetilde{\sigma}_y$. Both $E(\rho)$ and $C(\rho)$ range from 0 to 1, and are zero for unentangled states and 1 for maximally entangled states, as is desirable for a measure of entanglement.

\section{Quantum measurement}
\label{sec:measurement}
To fully specify an arbitrary state of a qubit, an infinite amount of information, in general, is needed: One has to announce the precise values of the complex numbers $\alpha$ and $\beta$ in Eq.~(\ref{eq:superpos}). However, due to the nature of quantum measurements, only a finite amount of information can ever be extracted from a qubit. Hence, the exact values of $\alpha$ and $\beta$ of an unknown qubit state can never be learned with certainty.

The formalism of quantum measurements defines how much information can be gained on an array of qubits, with a specific measurement scheme. It should be noted that, in quantum mechanics, measurements are the only way of acquiring knowledge on a previously unknown physical system. In the course of our analysis, we employ two kinds of measurements: \emph{projective} and \emph{positive operator-valued measure} (POVM) measurements.

\subsection{Projective measurements}
Projective measurements are described in any reasonable introduction to quantum mechanics. Therefore, only a short definition is presented here. A projective measurement has a corresponding observable $M$, which has a decomposition
\begin{equation}
M = \sum_m mP_m \; ,
\end{equation}
where $\{P_m\}$ are orthogonal projection operators to the eigenspaces with respective eigenvalues $\{m\}$. The eigenvalues are the possible outcomes upon measuring the observable $M$. If a state $|\psi\rangle$ is measured, the probability of outcome $m$ is
\begin{equation}
\label{eq:measprob}
p(m) = \langle\psi| P_m |\psi\rangle \; .
\end{equation}
Immediately after the measurement yielding $m$, the state of the system collapses to
\begin{equation}
\label{eq:measstate}
\frac{P_m |\psi\rangle}{\sqrt{p(m)}} \; .
\end{equation}
%That is, projective measurements do just what their name implies---project.

\subsection{Positive operator-valued measures}
Projective measurements are in general suitable for situations where there is interest in the evolution of the quantum system after the measurement. The converse is true for the POVM formalism which is more general than that of projective measurements, but still well suited for describing the probabilities of different outcomes.

Suppose there is a set of positive operators\footnote{A positive operator $A\!:\mathcal{H} \to \mathcal{H}$ is such that $\langle v| A |v\rangle$ is real and non-negative for any $|v\rangle \in \mathcal{H}$.} $\{E_m\}$ such that
\begin{equation}
\sum_m E_m = I \; .
\end{equation}
Then the set $\{E_m\}$ is a POVM, and the operators $E_m$ are known as the corresponding \emph{POVM elements}. The measurement that the POVM describes yields outcome $m$ for state $|\psi\rangle$ with probability
\begin{equation}
p(m) = \langle\psi|E_m|\psi\rangle \; .
\end{equation}

For example, any projective measurement can be described with a POVM: The POVM elements are the projectors of the projective measurement, $E_m = P_m$. In general however, POVM elements need not be projectors. Hence, POVMs clearly describe a larger class of measurements than projective measurements alone.

Another example, adopted from Ref.~\cite{nielsenchuang}, presents a measurement scheme with which one can possibly distinguish between states $|\varphi_1\rangle = |0\rangle$ and $|\varphi_2\rangle = \frac{1}{\sqrt{2}}(|0\rangle + |1\rangle)$, and never make an error of identifying state $|\varphi_1\rangle$ as $|\varphi_2\rangle$, or vice versa. Note that this characteristic cannot be achieved using only projective measurements. The POVM implementing this measurement scheme is
\begin{eqnarray}
E_1	& =	& \frac{\sqrt{2}}{1+\sqrt{2}} |1\rangle\langle1| \; , \\
E_2	& =	& \frac{\sqrt{2}}{2(1+\sqrt{2})} (|0\rangle - |1\rangle) (\langle0| - \langle1|) \; , \\
E_3	& =	& I - E_1 - E_2 \; .
\end{eqnarray}
Now, suppose we measure a given state using the POVM $\{E_1, E_2, E_3\}$, knowing that the state is either $|\varphi_1\rangle$ or $|\varphi_2\rangle$. If the measurement yields 1, we know that the given state was $|\varphi_2\rangle$. Similarly, result 2 implies that the state was $|\varphi_1\rangle$.
%However, result 3 is possible for both states. Hence, if the measurement result is 3, the given state could have been either of the two alternatives.

\section{Quantum cloning}
\label{sec:cloning}
\subsection{The no-cloning theorem}
\label{sec:no-cloning}
Unlike classical information, quantum information, in general, cannot be copied perfectly. This is due to the no-cloning property of quantum mechanics, stated as the \emph{no-cloning theorem}: It is not possible to produce perfect copies of an unknown quantum state. Applied to qubits, the theorem states that given a qubit in state $|Q\rangle = \alpha |0\rangle + \beta|1\rangle$, we cannot produce the state $|Q\rangle\otimes|Q\rangle$ without knowing the values of $\alpha$ and $\beta$. The theorem does not state that we cannot produce a number of similar states $|\Psi\rangle = |\psi\rangle\otimes|\psi\rangle\otimes\cdots\otimes|\psi\rangle$, only that given a system in state $|\psi\rangle$ with no additional information, we cannot create state $|\Psi\rangle$.

Let us give an elementary proof of the theorem, using \emph{reductio ad absurdum}. The evolution of any quantum system can be described as unitary by appending an auxiliary system, or \emph{ancilla}, to the original system. Now, suppose that a flawless copier exists. The copier has two slots, labelled $S_1$ and $S_2$, and an ancilla. The ancilla may be of any dimension. Slot $S_1$ occupies state $|\psi\rangle$ representing the source, i.e., the state we wish to reproduce. Slot $S_2$ is the target slot, which is to occupy the source state $|\psi\rangle$ after the copying process. Slot $S_2$ starts out in some initial state, denoted by $|s_0\rangle$. The initial state of the ancilla is $|\mathcal{A}_0\rangle$. That is, the initial state of the copying machine is
\begin{equation}
|\psi\rangle\otimes|s_0\rangle\otimes|\mathcal{A}_0\rangle \;.
\end{equation}
The copying process, which can be assumed unitary, and is also assumed to leave the source and target states and the ancilla unentangled, is denoted by $U_c$, and it clones the state in $S_1$ as \cite{scarani}
\begin{equation}
\label{eq:noclone2}
U_c (|\psi\rangle|s_0\rangle|\mathcal{A}_0\rangle) = |\psi\rangle|\psi\rangle|\mathcal{A}_\psi\rangle \; .
\end{equation}
The ancilla is left in state $|\mathcal{A}_\psi\rangle$ that possibly depends on $|\psi\rangle$. The same must hold for copying the state $|\phi\rangle \neq |\psi\rangle$:
\begin{equation}
U_c (|\phi\rangle|s_0\rangle|\mathcal{A}_0\rangle) = |\phi\rangle|\phi\rangle|\mathcal{A}_\phi\rangle \; ,
\end{equation}
and for the superposition state $|\Sigma\rangle = \frac{1}{\sqrt{2}}(|\phi\rangle + |\psi\rangle)$:
\begin{eqnarray}
U_c (|\Sigma\rangle|s_0\rangle|\mathcal{A}_0\rangle) & = & |\Sigma\rangle|\Sigma\rangle|\mathcal{A}_\Sigma\rangle \nonumber \\
																			 & = & \frac{1}{2}(|\phi\rangle + |\psi\rangle)(|\phi\rangle + |\psi\rangle)|\mathcal{A}_\Sigma\rangle \nonumber \\
																			 & = & \frac{1}{2}(|\phi\rangle|\phi\rangle + |\phi\rangle|\psi\rangle + |\psi\rangle|\phi\rangle + |\psi\rangle|\psi\rangle)|\mathcal{A}_\Sigma\rangle \label{eq:clonesigma} \; .
\end{eqnarray}
But because of linearity of quantum mechanics
\begin{eqnarray}
U_c (|\Sigma\rangle|s_0\rangle|\mathcal{A}_0\rangle) & = & \frac{1}{\sqrt{2}}\big( U_c (|\phi\rangle|s_0\rangle|\mathcal{A}_0\rangle) + U_c (|\psi\rangle|s_0\rangle|\mathcal{A}_0\rangle)\big) \nonumber \\
																			& = & \frac{1}{\sqrt{2}} \big(|\phi\rangle|\phi\rangle|\mathcal{A}_\phi\rangle + |\psi\rangle|\psi\rangle|\mathcal{A}_\psi\rangle\big) \; ,
\end{eqnarray}
which is clearly not compatible with Eq.~(\ref{eq:clonesigma}), even if there were no ancilla in the copier. Hence, the assumption of the existence of a perfect cloning machine must be false.

\subsection{Optimal cloning}
\label{sec:optimalclone}
We have shown that perfect cloning of quantum states is not possible. However, one can produce imperfect copies that are close to the original state with respect to some measure of cloning quality. Let us summarize the results on best possible discrete-system quantum cloning presented in Ref.~\cite{scarani}. Scarani \textit{et al.}~define $N \to M$ cloning of pure states as
\begin{equation}
(|\psi\rangle^{\otimes N})\otimes(|s_0\rangle^{\otimes(M-N)})\otimes|\mathcal{A}_0\rangle \stackrel{U}{\to} |\Psi\rangle \;
\end{equation}
where we continue the use of notation presented in Sec.~\ref{sec:no-cloning}, and $U$ is the unitary cloning process. The quality of the produced copies is measured as \emph{fidelity} $F$, which is defined
\begin{equation}
\label{eq:fidelity}
F_j := \langle\psi|\rho_j|\psi\rangle \;, \quad j\in\{1,...,M\} \;,
\end{equation}
where $\rho_j$ is the density operator of the partial state of clone $j$ in $|\Psi\rangle$. For a \emph{universal} copier, $F_j$ is independent of the source state $|\psi\rangle$. For a \emph{symmetric} copier, $F_j$ is independent of $j$. A copier is \emph{optimal} if it produces copies with maximal fidelity allowed by quantum physics, for a given source-state fidelity. The fidelity to be maximized can be the average fidelity $\bar{F} := \int_\mathcal{S} d\psi F(\psi)$, or the minimum of fidelities $F_{min} := \min_{\psi\in\mathcal{S}}F(\psi)$, where $\mathcal{S}$ is the set of source states.

The first quantum copying machine was presented by V.\ Bu\v zek and M.\ Hillery in 1996 \cite{buzek}. Their quantum copier is a universal, symmetric, optimal $1 \to 2$ copier capable of duplicating a qubit, and achieving a fidelity of $\frac{5}{6}$.
%This generalizes to the universal, symmetric $N \to M$ qubit cloner, presented first by N. Gisin and S. Massar \cite{gisinmassar}, whose optimal fidelity is
%\begin{equation}
%F^{N \to M}_{\mathrm{qubit}} = \frac{M N+M+N}{M(N+2)} \; .
%\end{equation}
This generalizes to the universal, symmetric, optimal $N \to M$ cloner for systems of any dimension $d$ (not only for qubits, where $d=2$), whose fidelity is
\begin{equation}
F^{N \to M}_d = \frac{N}{M} + \frac{(M-N)(N+1)}{M(N+d)} \; ,
\end{equation}
a result by R.\ F.\ Werner \cite{werner}.

In asymmetric cloning, the produced copies have different fidelities. For example, the universal, asymmetric, optimal $1 \to 2$ qubit-copying machine produces copies $A$ and $B$, whose fidelities are
\begin{equation}
F_A = 1-\frac{b^2}{2} \quad \mathrm{and} \quad F_B = 1-\frac{a^2}{2} \; ,
\end{equation}
where the real parameters $a$ and $b$ satisfy $a^2+b^2+ab=1$. This result was found independently by several groups \cite{niugriffiths,cerf,scarani}.
%C.-S.~Niu and R.~B.~Griffiths \cite{niugriffiths}, N.~J.~Cerf \cite{cerf}, and V. Bu\v zek \textit{et al.} \cite{scarani}.

Again, the above mentioned qubit copier generalizes to the universal, asymmetric, optimal $1 \to 2$ cloner capable of copying systems of dimension $d$, described in Refs.~\cite{cerf2, braunstein}. Optimality was, however, proven later \cite{cloneopt2,cloneopt3}. The fidelities of the output systems of the cloner are
\begin{equation}
F_A = 1-\frac{d-1}{d}b^2 \quad \mathrm{and} \quad F_B = 1-\frac{d-1}{d}a^2 \; ,
\end{equation}
where the real parameters $a$ and $b$ now satisfy $a^2+b^2+\frac{ab}{d}=1$.

Finally, we review the fidelity of a non-universal, i.e., state-dependent, $1 \to 2$ qubit cloner, which is in addition \emph{phase covariant}. In this context, phase covariance means that the cloner can clone, at best, states of the form
\begin{equation}
\label{eq:phasecovariant}
|\Psi(\varphi)\rangle = \frac{1}{\sqrt{2}}\big(|0\rangle + e^{i\varphi}|1\rangle\big) \; .
\end{equation}
The cloner, discussed in Ref.~\cite{niugriffiths99}, transforms input states as
\begin{eqnarray}
|0\rangle|0\rangle	& \to	& |0\rangle|0\rangle \nonumber \\
|1\rangle|0\rangle	& \to	& \cos(\eta)|1\rangle|0\rangle + \sin(\eta)|0\rangle|1\rangle \; , \label{eq:phcovcloner}
\end{eqnarray}
where $\eta \in [0,\frac{\pi}{2}]$. Applying this to the state in Eq.~(\ref{eq:phasecovariant}) yields
\begin{equation}
\label{eq:phcovtrans}
|\Psi(\varphi)\rangle|0\rangle \to \frac{1}{\sqrt{2}}\big(|0\rangle|0\rangle + \cos(\eta) e^{i\varphi} |1\rangle|0\rangle + \sin(\eta) e^{i\varphi} |0\rangle|1\rangle \big) \; ,
\end{equation}
from which one obtains the fidelities
\begin{equation}
\label{eq:phcovfids}
F_A = \frac{1}{2}(1+\cos\eta) \quad \mathrm{and} \quad F_B = \frac{1}{2}(1+\sin\eta) \; .
\end{equation}
Note that these fidelities are independent of the phase factor $\varphi$.

%% file: qkd.tex
\chapter{Quantum Key Distribution}
\label{ch:qkd}
Quantum key distribution (QKD) refers to any scheme that allows two distant parties to securely establish a shared, secret string of bits, and in which the security is based on the laws of quantum physics. C.\ H.\ Bennett and G.\ Brassard were the first to propose a practical protocol for QKD \cite{bb84}, known as the BB84 protocol. Section \ref{sec:bb84} describes the protocol in detail. The first experimental realization of the protocol is described in Ref.~\cite{bb84implemented}. During the last two decades, the scientific community has introduced an overwhelming amount of modifications to the BB84 protocol, as well as entirely new QKD protocols \cite{mod2state,mod6state,modbasesprob,modSARG,modEPR,modqudits,modsothers1,modsothers2,modsothers3,modsothers3b,modsothers4,modsothers5,modsothers6,modsothers7,modsothers8,modsothers9,modsothers10,modsothers11,modsothers12,modsothers13,modsothers14,modsothers15,modsothers16,modsothers17,modsothers18,modsothers19,modsothers20,modsothers21,modsothers22,modsothers23,modsothers24}. Some of these are briefly reviewed in Sec.~\ref{sec:qkdother}.

Let us introduce the customary terminology of quantum cryptography. The inititator of the communication is \emph{Alice}, and the party to whom Alice wants to send her messages is \emph{Bob}. The all-purpose malevolent party who wishes to spy on Alice's and Bob's communication is \emph{Eve}, the eavesdropper. If Alice or Bob co-operate with a non-malicious third party, he is called \emph{Charlie}. A \emph{channel} allows Alice and Bob to send data to each other. A channel can be one-way or two-way. A \emph{classical channel} transmits bits, and a \emph{quantum channel} transmits qubits. A \emph{public channel} is one that anyone can listen to and to which anyone can send messages. If Alice and Bob use an \emph{authenticated channel}, Eve cannot send messages such that they would appear to Bob to be from Alice, or vice versa. When considering the security of a particular cryptographic scheme, Eve is granted various capabilities. The set of Eve's capabilities together with her actions during the execution of a protocol defines an \emph{attack}. Section \ref{sec:bb84attacks} describes attacks against the BB84 protocol. Section \ref{sec:qkdother} summarizes attacks against other protocols.

\section{BB84 protocol}
\label{sec:bb84}

\begin{figure}[hbtp]
\begin{center}
\includegraphics[width=0.65\textwidth]{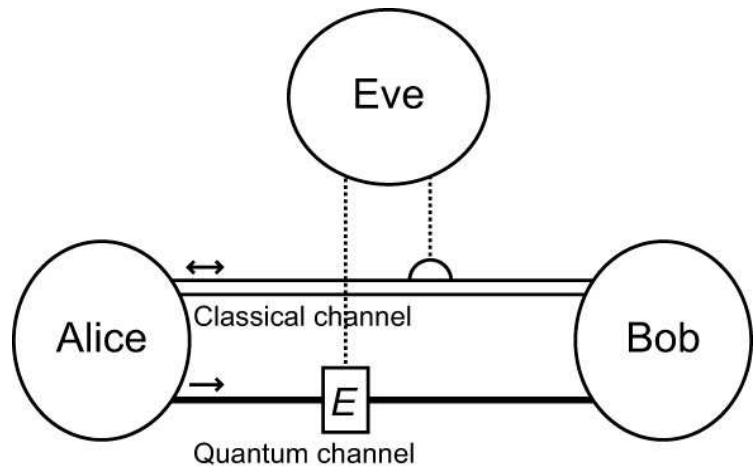}
\caption{Setup of the BB84 quantum key distribution protocol. Alice and Bob use a two-way classical channel and a one-way quantum channel. Eve has total control over the quantum channel, whereas she can only listen to the classical channel.}
\label{fig:bb84setup}
\end{center}
\end{figure}

Figure \ref{fig:bb84setup} shows a schematic illustration of the BB84 protocol. Alice and Bob are in possession of a two-way public authenticated classical channel. They also have a one-way public quantum channel, which allows Alice to send individual qubits to Bob. Eve is allowed total control of the quantum channel. That is, she can delete and insert transmissions, as well as alter them in any way that is not forbidden by the laws of quantum mechanics. Eve's interaction with the channel is denoted by $E$. In addition, Eve is assumed to listen in on every transmission on the classical channel. In the following description, we first assume no participation on Eve's behalf, and postpone the discussion of the effects introduced by Eve's interference to Sec.~\ref{sec:bb84attacks}. Likewise, the discussion of an imperfect quantum channel is postponed to Sec.~\ref{sec:nonideal}, and for now, both channels are assumed ideal, i.e., error-free. There is no Charlie in this protocol, that is, Alice and Bob do not have to rely on any third party to complete the key distribution.

\subsection{Transmission}
When Alice and Bob have decided, e.g., using the classical channel, to initiate the protocol, Alice begins the transmission of individual particles on the quantum channel. In the original paper, these particles are photons \cite{bb84}, and hence we will describe the protocol with photon transmission. However, any quantum system qualifying as a flying qubit\footnote{See Sec.~\ref{sec:realqubit}.} with two maximally conjugate bases would serve. Each of Alice's photons randomly occupy one of four possible states. Each state corresponds to a direction of linear polarization of the photon.

The choice of the state of each photon can be considered to consist of two binary random variables. The actual physical systems corresponding to these random variables are in Alice's possession. The first random variable $A^\prime$ represents the candidate for the bit value that Alice is trying to send to Bob. It is essential that $p(A^\prime=0) = p(A^\prime=1) = \frac{1}{2}$. Alice records for herself the outcomes of $A^\prime$. The second random variable $P_\mathrm{A}$ determines in which basis the output of $A^\prime$ is transmitted. For $P_\mathrm{A}$ as well, we have to have $p(P_\mathrm{A}=z) = p(P_\mathrm{A}=x) = \frac{1}{2}$.

If $P_\mathrm{A}=z$, Alice encodes the outcome 0 of $A^\prime$ as vertical polarization of the photon, and the outcome 1 as horizontal polarization of the photon. The vertical and horizontal polarization states are denoted by $|\!\updownarrow\,\rangle$ and $|\!\leftrightarrow\rangle$, respectively. Photons transmitted in either of these states are said to be sent in the $\oplus$ basis. If $P_\mathrm{A}=x$, Alice uses a diagonal basis: The outcome 0 of $A^\prime$ is encoded as the 45$^{\circ}$ rotated linear polarization, vertical being the non-rotated direction of polarization, and the outcome 1 of $A^\prime$ as the 135$^{\circ}$ rotated linear polarization. The respective states are denoted by $|\!\nearrow\!\!\!\!\!\!\swarrow\,\rangle$ and $|\!\searrow\!\!\!\!\!\!\nwarrow\,\rangle$. This is known as the $\otimes$ basis.

The photon transmission states and bases can be equivalently described using the spin formalism of quantum mechanics. If $P_\mathrm{A} = z$, Alice transmits in the eigenbasis of the Pauli spin matrix $\sigma_z$: The outcome 0 of $A^\prime$ is sent as $|0\rangle$, or equivalently as the spin-up state $|\!\uparrow\rangle$, and the outcome 1 as $|1\rangle$, equivalent to the spin-down state $|\!\downarrow\rangle$. If $P_\mathrm{A} = x$, Alice transmits in the eigenbasis of the Pauli spin matrix $\sigma_x$: The outcome 0 of $A^\prime$ is sent as $|+\rangle = \frac{1}{\sqrt{2}}(|0\rangle + |1\rangle)$, and the outcome 1 as $|-\rangle = \frac{1}{\sqrt{2}}(|0\rangle - |1\rangle)$. In summary, $|0\rangle = |\!\updownarrow\,\rangle$, $|1\rangle = |\!\leftrightarrow\rangle$, $|+\rangle = |\!\nearrow\!\!\!\!\!\!\swarrow\,\rangle$,  $|-\rangle = |\!\searrow\!\!\!\!\!\!\nwarrow\,\rangle$, and basis $\oplus$ corresponds to the $\sigma_z$ eigenbasis and basis $\otimes$ to the $\sigma_x$ eigenbasis---hence the labels $z$ and $x$ for the outcomes of $P_\mathrm{A}$. The $z$ and $x$ bases are maximally conjugate in the sense that for any pair of states chosen from different bases, the square modulus of the inner product is $|\langle0|+\rangle|^2 = |\langle0|-\rangle|^2 = \ldots = \frac{1}{2}$.

\subsection{Measurement}
Upon reception, Bob measures the polarization of each arriving photon. Bob's particular measurement is defined by a random variable $P_\mathrm{B}$, whose physical system Bob is in total control of. The random variable $P_\mathrm{B}$ is identical to $P_\mathrm{A}$ in the sense that $p(P_\mathrm{B}=z) = p(P_\mathrm{B}=x) = \frac{1}{2}$ but totally independent of $P_\mathrm{A}$. If $P_\mathrm{A}$ and $P_\mathrm{B}$ were to depend on each other in some way, Alice and Bob would have to exchange information as $P_\mathrm{A}$ and $P_\mathrm{B}$ assume their values. This cannot be allowed, as it would severely compromise the security of the protocol. Therefore, it is required that $P_\mathrm{A}$ and $P_\mathrm{B}$ are independent.

Bob chooses the basis for his measurement of the direction of polarization in the same way that Alice chooses her basis of transmission. If $P_\mathrm{B} = z$, Bob measures in $\sigma_z$ eigenbasis, and if $P_\mathrm{B} = x$, he measures in $\sigma_x$ eigenbasis. Bob uses a projective measurement for each basis choice. For the $z$ basis, the measurement projectors are $P_0^z = |0\rangle\langle0|$ and $P_1^z = |1\rangle\langle1|$. For the $x$ basis, Bob uses $P_0^x = |+\rangle\langle+|$ and $P_1^x = |-\rangle\langle-|$. For example, in the $x$ basis, when $A^\prime=0$, Bob recovers this with probability
\begin{equation}
p(0) = \langle+| P_0^x |+\rangle = \langle+|+\rangle\langle+|+\rangle = |\langle+|+\rangle|^2 = 1 \; .
\end{equation}
We observe that whenever the bases chosen by Alice and Bob coincide, Bob exactly recovers the value of Alice's random variable $A^\prime$. This happens with probability
\begin{eqnarray}
\label{eq:coinprob}
&		& p(P_\mathrm{A} = P_\mathrm{B} = z \mathrm{~OR~} P_\mathrm{A} = P_\mathrm{B} = x)	\nonumber	\\
& =	& p(P_\mathrm{A} = z, P_\mathrm{B} = z) + p(P_\mathrm{A} = x, P_\mathrm{B} = x)				\nonumber	\\
& =	& p(P_\mathrm{A} = z)p(P_\mathrm{B} = z) + p(P_\mathrm{A} = x)p(P_\mathrm{B} = x)			\nonumber	\\
& =	& \frac{1}{2} \; ,																													
\end{eqnarray}
where we have used Eq.~(\ref{eq:probor}), the fact that only one basis is defined for Alice and Bob at a time, and the independence of $P_\mathrm{A}$ and $P_\mathrm{B}$.

From Eq.~(\ref{eq:coinprob}), it follows that Bob uses a basis incompatible with Alice's basis with probability $\frac{1}{2}$. When this happens, Bob cannot recover the value of $A^\prime$. For instance, if Alice has transmitted $|+\rangle$ and Bob measures this in the wrong basis,
\begin{equation}
p(0) = \langle+|P_0^z|+\rangle = \frac{1}{2}(\langle0| + \langle1|)|0\rangle\langle0|(|0\rangle + |1\rangle) = \frac{1}{2}\langle0|0\rangle = \frac{1}{2} \; .
\end{equation}
In fact, the same probability is obtained for each result and for each photon state given that Bob chooses the wrong basis for his measurement. That is, when the bases are not compatible, Bob gets the two possible results with equal probability.

\subsection{Basis reconciliation}

After each measurement, Bob interprets results $|0\rangle$ and $|+\rangle$ as 0, and results $|1\rangle$ and $|-\rangle$ as 1, and records this interpretation for himself. Bob's sequence of interpretations from the individual polarization measurements is known as the \emph{raw key}. Because Alice and Bob choose the same bases with probability $\frac{1}{2}$, in the limit of a long key, only half of the bits in Bob's raw key are definitely the same as the output of $A^\prime$ recorded by Alice. For the measurements where Alice's and Bob's bases do not coincide, the result is, by chance, correct half the time, so on average half of this half of the raw-key bits agree. Since this is as good as Bob having simply guessed the values without any measurement, these bits have no value in this protocol, and Bob should discard them.

To be able to decide which bits to discard, Bob sends the sequence of his basis choices, i.e., outcomes of $P_\mathrm{B}$, to Alice through the classical channel. Alice replies, through the classical channel, with her basis choices, i.e., outcomes of $P_\mathrm{A}$. This is called basis reconciliation: Alice and Bob compare their basis-choice sequences, and discard all the bits where Bob used the wrong basis. That is, Alice and Bob keep only those bits for which their bases happened to coincide. What is left is an error-free shared string of bits known as the \emph{sifted key}. Thus the protocol has achieved its goal. Table \ref{tab:bb84ex} presents an example of the use of this protocol.

\begin{table}[hbt]
\begin{center}
\caption{A brief example of the BB84 QKD protocol. This example assumes error-free channels and no interference by Eve. Alice sends the photons in the state determined by her key-candidate and transmission-basis random variables. Bob measures the state of the photons in a randomly chosen basis, shown on line 5, and records his results (line 6). After the photon transmission is complete, Alice and Bob compare their basis choices and discard the bits for which their bases did not match. The remaining bits are shown on line 8. Finally, Alice and Bob choose randomly which bits to compare publicly (line 9), in order to estimate the quantum bit error rate of the transmission, described in Sec.~\ref{sec:nonideal}.}
\label{tab:bb84ex}
\vspace{2mm}
\begin{tabular}[c]{llcccccccccccc}
\hline
\hline
1&	Transmission								&	1$^{\mathrm{st}}$	&	2$^{\mathrm{nd}}$	&	3$^{\mathrm{rd}}$	&	4$^{\mathrm{th}}$	&	5$^{\mathrm{th}}$	&	6$^{\mathrm{th}}$	&	7$^{\mathrm{th}}$	&	8$^{\mathrm{th}}$	&	9$^{\mathrm{th}}$	&	10$^{\mathrm{th}}$ \\
\hline
2&	$A^\prime$									&	1							&	0							&	1							&	1							&	0							&	1							&	1							&	0							&	0							&	1							\\
3&	$P_\mathrm{A}$							&	$z$						&	$x$						&	$x$						&	$z$						&	$z$						&	$z$						&	$x$						&	$z$						&	$x$						&	$z$						\\
4&	Alice sends									&	$|1\rangle$			&	$|+\rangle$			&	$|-\rangle$				&	$|1\rangle$			&	$|0\rangle$			&	$|1\rangle$			&	$|-\rangle$				&	$|0\rangle$			&	$|+\rangle$			&	$|1\rangle$			\\
\hline
5&	$P_\mathrm{B}$							&	$x$						&	$z$						&	$x$						&	$x$						&	$z$						&	$z$						&	$z$						&	$z$						&	$x$						&	$x$						\\
6&	Raw key										&	0							&	1							&	1							&	0							&	0							&	1							&	1							&	0							&	0							&	1							\\
\hline
7&	$P_\mathrm{A} = P_\mathrm{B}$	&	no						&	no						&	yes						&	no						&	yes						&	yes						&	no						&	yes						&	yes						&	no						\\
8&	Sifted key										&								&								&	1							&								&	0							&	1							&								&	0							&	0							&								\\
\hline
9&	Error estim.									&								&								&	no						&								&	no						&	yes						&								&	no						&	yes						&								\\
10&	Key												&								&								&	1							&								&	0							&								&								&	0							&								&								\\
\hline
\hline
\end{tabular}
\end{center}
\end{table}

For later purposes, it is convinient to model also the result of Bob's measurement, given that he used the same basis as Alice, as a random variable. Thus the outcomes of this random variable $B$ determine the bits in Bob's sifted key. We define a similar random variable for Alice: Outcomes of random variable $A$ determine the bit values in Alice's sifted key, i.e., in the bit sequence in Alice's possession after basis reconciliation. The outcomes of $A$ are a subset of the outcomes of $A^\prime$. Bob's random variable $B$ is, of course, highly dependent on $A$. For now, we state
\begin{eqnarray}
&&p(B=0|A=0) = p(B=1|A=1) = 1 \label{eq:errfree1}	\\
&&p(B=0|A=1) = p(B=1|A=0) = 0 \label{eq:errfree2}	\;.
\end{eqnarray}
These equations cease to hold if the assumption of non-interfering Eve is relaxed, or if the quantum channel is allowed a finite error rate.

\subsection{Issues introduced by non-ideal equipment}
\label{sec:nonideal}
The technology Alice and Bob use to implement a QKD protocol is never perfect. This section reviews the most important implications of the use of non-ideal equipment for the BB84 protocol. Firstly, the quantum channel conveying quantum states from Alice to Bob is not perfect. Alice's transmission may be totally lost, or its polarization may be randomly rotated by a small angle. A lossy quantum channel poses no fundamental problem for Alice and Bob, as they can agree that Alice uses the classical channel to tell Bob when she transmits, and that Bob tells Alice which transmissions were succesfully received. A quantum channel that randomly rotates the polarization of the photons causes errors to the sifted key, which means that Eqs.~(\ref{eq:errfree1}) and (\ref{eq:errfree2}) do not hold anymore. The probability that Bob's measurement yields an incorrect result, even if he uses the same basis as Alice, is coined \emph{quantum bit error rate} (QBER). That is, assuming an error process independent of the direction of polarization, $p(B=0|A=1) = p(B=1|A=0) = \mathrm{QBER}$. Alice and Bob can obtain an estimate of the QBER by publicly comparing the values of a fraction of their respective sifted keys. Subsequently, they have to discard the bits whose values have been announced in public. Table \ref{tab:bb84ex} presents an example of this step. Working BB84-based QKD schemes have been reported with a QBER ranging from 1.0\% to 10.2\% \cite{plugnplay,qber102,qber0102,qber034}. Errors in the sifted key can be corrected using classical error correction procedures with communication over the classical channel, described in Sec.~\ref{sec:errcorr}.

Secondly, Bob's detectors are imperfect: Sometimes a detector fails to register a photon, and sometimes it incorrectly reports to have received a photon when in reality no transmission was received. Reports of the latter type are known as \emph{dark counts}. Both of these effects can be counteracted with the same technique that was used to deal with a lossy channel: Alice and Bob declare their transmissions and receptions over the classical channel. Subsequently, Alice discards bits lost in the quantum channel or Bob's detector, and Bob discards excess bits created by dark counts. The rare event where a photon sent by Alice was lost in the channel, but Bob still observes a reception because of a dark count, contributes to the QBER. For instance, the following figures have been observed achievable in QKD experiments. A. Muller \textit{et al.}~have implemented the original BB84 protocol using photon polarization, reporting a detector efficiency of 0.2\% and a dark count rate of 700 s$^{-1}$, with 1.1$\cdot10^6$ s$^{-1}$ transmission rate at Alice's end \cite{qber034}, T. Hirano \textit{et al.}~have implemented a BB84 variant with detector efficiencies near 80\% \cite{qber0102}, and L. P. Lamoureux \textit{et al.}~report a 10.5\% detector efficiency in a quantum coin-tossing protocol \cite{lamoureux}.

There are also several technical issues affecting the security of an implementation. One of the most serious problems is due to the fact that the BB84 protocol assumes that Alice can send individual photons to Bob. In reality, however, reliably creating transmissions containing exactly one photon is very difficult. Usually, single-photon pulses for QKD are created with an attenuated laser, for which the number of photons per pulse is a Poisson-distributed random variable \cite{bouwmeester}. Thus some of the pulses do not contain a photon at all, and some pulses contain one, two, or even more photons. Pulses containing more than one photon compromise the security of the protocol, since Eve can mount a specific attack based on multi-photon transmissions \cite{beamsplitter}. Therefore, the probability of transmitting more than one photon per transmission should be very small. Consequently, the probability of transmitting at least one photon tends to be quite small, as well. In Ref.~\cite{bouwmeester}, A.\ Ekert \textit{et al.}~present the following figures: The usual source used in QKD emits, on average, 0.1 photons per pulse, and 5\% of the pulses that contain at least one photon, contain more than one photon. The authors anticipate that these figures will improve as technology advances. For instance, B.\ Darqui\'e \textit{et al.} reported in 2005 an experiment with a triggered source emitting single right-circularly polarized\footnote{Circular as well as linear polarization states can be used directly in BB84 \cite{gisin}.} photons \cite{darquie}. Pulses from this source contain one photon with probability 0.981, and two photons with probability 0.019.

The implementation of the quantum channel of the protocol, including the detectors of Alice and Bob, may offer Eve the possibility of a \emph{trojan horse} attack. In this type of attack, for example, Eve sends pulses of light to the quantum channel and observes the pattern of light reflected back from Alice's and Bob's equipment \cite{gisin}. This way, Eve can, in principle, acquire information on the bases used by Alice and Bob, on the last value of Alice's key candidate variable $A$, or on the result of Bob's last measurement. Secure measures exist to thwart the trojan horse attack \cite{sciencelo}.
%korjaus 28.12. 

\subsection{Attacks}
\label{sec:bb84attacks}
In this section, we discuss attacks on the ideal BB84 protocol. That is, we abide by the framework presented in Fig.~\ref{fig:bb84setup} and assume that Alice and Bob have taken all necessary precautions to counteract any security threats introduced by non-ideal equipment. Furthermore, Eve does not have access to Alice's or Bob's office, e.g., she cannot use a telescope to watch Bob's display. To recapitulate, Eve can only:
\begin{itemize}
\item[i)] Freely tamper with the quantum channel. 
\item[ii)] Listen in on everything that is transmitted on the classical channel.
\end{itemize}

A characteristic feature of quantum key distribution schemes is that any known method of eavesdropping inevitably causes errors to the quantum transmission, increasing the QBER. The errors allow Alice and Bob to detect Eve's interference and to obtain an estimate on Eve's maximal information about the key. In BB84, the QBER is the only, albeit guaranteed, indicator of Eve's interference. As described in Sec.~\ref{sec:nonideal}, an error estimate is obtained by publicly comparing random bits in the sifted key. The accuracy of the estimate can be made arbitrarily high by increasing the number of compared bit values. An example of the estimation is included in Table \ref{tab:bb84ex}. Alice and Bob have no way of resolving which errors are due to an imperfect quantum channel and which are due to Eve's actions, and they therefore safely assume that the estimated QBER is in its entirety due to Eve. After all, Eve could have, in principle, replaced most of the noisy quantum channel with a less noisy one. Alice and Bob correct errors of either origin using a classical error correction protocol.

Considering what Eve should do to gain knowledge on Alice and Bob's key, perhaps the first tactics that would come to one's mind is that Eve would capture each of Alice's qubit transmissions, prepare a copy of each for herself, and send another copy to Bob. Note that Eve has to send something to Bob, to allow him to continue the protocol---otherwise the transmission would never result in a key. Eve could keep her copies intact until Alice and Bob publicly announce their basis choices, and measure each transmission in the correct basis. She would then obtain a flawless copy of the key that Alice and Bob established without them knowing of this at all. However, this is where quantum physics steps in: According to the no-cloning theorem (Sec.~\ref{sec:no-cloning}), Eve cannot make perfect copies of all the states transmitted in the protocol. Therefore, this type of attack is simply not possible. However, if Eve settles for flawed copies, the attack is feasible and considerable. This imperfect cloning attack is equivalent to an \emph{incoherent attack}, discussed below.

%assumptions: isolation, only interesting attacks consdired, i.e. alice is not allowed to 'look over bob's shoulder', P_A and P_B are (P)RNGs
\subsubsection{Intercept-resend}
In the intercept-resend (IR) attack, Eve individually intercepts each qubit sent by Alice, measures the qubit state, and resends to Bob a qubit in the state corresponding to her measurement result. Eve performs her measurements exactly like Bob: For each qubit, she chooses at random between the two measurement bases: eigenbasis of $\sigma_z$ or eigenbasis of $\sigma_x$. Alternatively, Eve can use the same basis every time. This does not affect the analysis, since Alice always transmits in a randomly chosen basis. If Eve uses the $z$ basis in a measurement, result 0 means that Eve sends $|0\rangle$, and result 1 that she sends $|1\rangle$ to Bob. If Eve's measurement basis is $x$, she resends $|+\rangle$ if the result is 0, and $|-\rangle$ if the result is 1. Note that, on average, Eve inevitably chooses the wrong basis with probability $\frac{1}{2}$. Thus, Eve's interference increases the QBER, based on which Alice and Bob estimate Eve's maximal information on Alice's sifted key, i.e., the outcomes of the random variable $A$.

Let us calculate exactly how much information, on average, the IR attack maximally provides Eve as a function of QBER. To verify that Eve indeed chooses the wrong basis with probability half, consider the two cases: Eve uses the same basis every time, or Eve chooses her basis randomly and uniformly. In the first case, because Alice's transmission basis is half the time $z$ and half the time $x$, either fixed basis leads to Eve's choice being wrong on average half the time. As for the second case, let $P_\mathrm{E}$ denote Eve's measurement basis: $p(P_\mathrm{E}=z) = p(P_\mathrm{E}=x) = \frac{1}{2}$. The probability that Eve's choice is compatible with Alice's is
\begin{eqnarray}
\label{eq:evewrong}
&		& p(P_\mathrm{A} = P_\mathrm{E} = z \mathrm{~OR~} P_\mathrm{A} = P_\mathrm{E} = x)	\nonumber	\\
& =	& p(P_\mathrm{A} = z, P_\mathrm{E} = z) + p(P_\mathrm{A} = x, P_\mathrm{E} = x)				\nonumber	\\
& =	& p(P_\mathrm{A} = z)p(P_\mathrm{E} = z) + p(P_\mathrm{A} = x)p(P_\mathrm{E} = x)			\nonumber	\\
& =	& \frac{1}{2} \; ,																													
\end{eqnarray}
exactly like in Eq.~(\ref{eq:coinprob}), and thus the choice is wrong half the time. Of course, Eve could favor, say, the $z$ basis so that $p(P_\mathrm{E}=z) = 1 - p(P_\mathrm{E}=x) > \frac{1}{2}$. This strategy does not change the probability obtained above, since it is equivalent to using a fixed choice some of the time, and a random choice with uniform probabilities some of the time. The probability of choosing wrong is $\frac{1}{2}$ for both and thus Eve cannot increase, or decrease, the probability of her basis being the same as Alice's basis.

The knowledge that Eve has on Alice's bit sequence after basis reconciliation, i.e., outcomes of $A$, is quantified as the mutual information $I(A,E)$, where $E$ is a random variable denoting the outcome of each of Eve's measurements. However, $E$ only models measurements which correspond to transmissions that contribute to the sifted key. This is sensible because the basis reconciliation phase is public and thus Eve knows which bits Alice and Bob discard. According to Eq.~(\ref{eq:mi}),
\begin{equation}
\label{eq:aeinfo}
I(A,E) = H(A) + H(E) - H(A,E) \; .
\end{equation}

The entropy of Alice's random variable $A$ is
\begin{equation}
H(A) = -\sum_{a=0}^{1} p(a) \log p(a) = H_{\mathrm{bin}}\Big(\frac{1}{2}\Big) = 1 \; .
\end{equation}
To be able to express $H(E)$, we need to calculate the probability distribution of $E$, i.e., values $p(E=0)$ and $p(E=1)$. Since $p(E=1) = 1 - p(E=0)$, it is sufficient to determine $p(E=0)$. Eve can obtain the result $E=0$ in two mutually exclusive cases: Eve has the wrong basis, or Eve has the correct basis, compared to Alice's basis. Let $B_{\mathrm{w}}$ and $B_{\mathrm{c}}$ denote these events, respectively. In accordance with the law of total probability, Eq.~(\ref{eq:totalprob}), we have
\begin{eqnarray}
p(E=0)	& =	& p(E=0|B_{\mathrm{w}}) p(B_{\mathrm{w}}) + p(E=0|B_{\mathrm{c}}) p(B_{\mathrm{c}}) \nonumber \\
			& =	& \frac{1}{2} \left[p(E=0|B_{\mathrm{w}}) + p(E=0|B_{\mathrm{c}})\right] \; .
\end{eqnarray}

The case where Eve chooses the correct basis is straightforward to analyze: Alice's transmission is in a state that already lies in the subspace of Eve's measurement projectors. For example,
\begin{equation}
p(E=0, P_\mathrm{E}=x | A=0, P_\mathrm{A}=x) = \langle+|P_0^x|+\rangle = \langle+|+\rangle\langle+|+\rangle = 1 \; .
\end{equation}
Hence, Eve always correctly obtains the outcome of $A$, and we get
\begin{eqnarray}
p(E=0 | B_{\mathrm{c}})	& \stackrel{(\ref{eq:probsum})}{=} & \sum_a p(E=0, a | B_{\mathrm{c}}) \nonumber\\
									& \stackrel{(\ref{eq:cond1})}{=}	& \sum_a p(E=0 | a, B_{\mathrm{c}}) p(a|B_{\mathrm{c}}) \nonumber\\
									& =											& \sum_a p(E=0 | a, B_{\mathrm{c}}) p(a) \nonumber\\
									& =											& \frac{1}{2} \left[p(E=0 | A=0, B_{\mathrm{c}}) + p(E=0 | A=1, B_{\mathrm{c}}) \right] \nonumber\\
									& =											& \frac{1}{2} \left(1+0\right) \nonumber\\
									& =											& \frac{1}{2} \; ,
\end{eqnarray}
where $a$ is the outcome of $A$, and we have used the fact that Alice's and Eve's basis choices are independent of $A$.

As for Eve's wrong choice of basis, the incorrect choice can be made in exactly two ways: by choosing the $z$ basis, or by choosing the $x$ basis. Furthermore, in each of these cases, Alice may have sent either 0 or 1. These four cases each yield probability $\frac{1}{2}$ for $E=0$, which can be observed by calculating
\begin{eqnarray}
p(E=0, P_\mathrm{E}=z | A=0, P_\mathrm{A}=x)	= \langle+|P_0^z|+\rangle = \langle+|0\rangle\langle0|+\rangle = |\langle+|0\rangle|^2 = \frac{1}{2} & & \nonumber\\
																		&		& \\
p(E=0, P_\mathrm{E}=x | A=0, P_\mathrm{A}=z)	= \langle0|P_0^x|0\rangle = \langle0|+\rangle\langle+|0\rangle = |\langle+|0\rangle|^2 = \frac{1}{2} \; ,& &  \nonumber\\
																		&		& 
\end{eqnarray}
and similarly for the cases where $A=1$. Thus, $p(E=0|B_{\mathrm{w}}) = \frac{1}{2}$, and we have $p(E=0) = p(E=1) = \frac{1}{2}$ or in short $p(e) = \frac{1}{2}$, from which we finally obtain the entropy of Eve's measurement outcome
\begin{equation}
H(E) = -\sum_{e=0}^{1} p(e) \log p(e) = H_{\mathrm{bin}}\Big(\frac{1}{2}\Big) = 1 \; .
\end{equation}

%korjaus 27.12. relating -> related
To calculate the third term in Eq.~(\ref{eq:aeinfo}), $H(A,E)$, we treat the cases of correct and incorrect basis separately, and use their average as the joint entropy $H(A,E)$. The use of an average is justified by noting that all the quantities related to Eve's knowledge about the key are averaged over a large set of transmissions from Alice to Bob. That is, we are dealing with probabilities. Using the definitions of joint entropy, Eq.~(\ref{eq:jointentr}), and conditional probability, Eq.~(\ref{eq:condprob}), we expand
\begin{eqnarray}
H(A,E)	& =	& -\sum_{a,e} p(a,e) \log p(a,e) \nonumber \\
			& =	& -\sum_{a,e} p(e|a)p(a) \log [p(e|a)p(a)] \nonumber \\
			& =	& -\frac{1}{2} \sum_{a,e} p(e|a) \log \Big[\frac{1}{2} p(e|a) \Big] \nonumber \\
			& =	& \frac{1}{2} \sum_{a,e} p(e|a) - \frac{1}{2} \sum_{a,e} p(e|a) \log p(e|a) \; .
\end{eqnarray}
We have already calculated the probabilities $p(e|a)$. If Eve chooses her basis correctly, $p(0|0) = p(1|1) = 1$ and $p(0|1) = p(1|0) = 0$. Thus the joint entropy in the case of correct basis is 1. When Eve chooses the wrong basis, $p(0|0) = p(0|1) = p(1|0) = p(1|1) = \frac{1}{2}$. Thus the joint entropy for an incorrect basis choice is 2. Because Eve's basis choice is correct on average half the time, $H(A,E) = \frac{3}{2}$. Applying the results to Eq.~(\ref{eq:aeinfo}) gives $I(A,E) = \frac{1}{2}$. That is, Eve gains 0.5 bits of information per bit in the sifted key.

The results are very intuitive. When Eve's basis is correct, $E$ gives exactly the same information as $A$ without error. When Eve's basis is incorrect, she gets results that are totally random, and $E$ conveys no information on $A$. The correct basis is used with probability $\frac{1}{2}$, i.e., half the time in a large set of interceptions. Therefore, Eve gets half of the bits in Alice's sifted key, and the rest is random noise.

There is an alternative and somewhat simpler formulation for Eve's knowledge on the sifted key, which we will later use in our analysis. One defines a composite random variable $\widetilde{A}$ for the joint outcome of $A$ and $P_\mathrm{A}$. That is, $\widetilde{A}$ describes the quantum state of Alice's transmission, and assumes its values $\tilde{a}$ from the set $\{0,1,+,-\}$ with uniform probability $p(\tilde{a}) = \frac{1}{4}$. In the following, we prove that calculating Eve's information on the transmission state $\widetilde{A}$ yields exactly the same result as calculating her information on $A$, i.e., $I(\widetilde{A},E) = I(A,E)$.

The mutual information of $\widetilde{A}$ and $E$ is
\begin{equation}
\label{eq:itildeae}
I(\widetilde{A},E) = H(\widetilde{A}) + H(E) - H(\widetilde{A},E) \; ,
\end{equation}
where $H(E) = 1$, as before, and
\begin{eqnarray}
H(\widetilde{A})	& =	& -\sum_{\tilde{a}=0}^{-} p(\tilde{a}) \log p(\tilde{a}) \nonumber\\
						& =	& -\sum_{\tilde{a}=0}^{1} p(\tilde{a}) \log p(\tilde{a}) -\sum_{\tilde{a}=+}^{-} p(\tilde{a}) \log p(\tilde{a}) \nonumber\\
\label{eq:etildea}& =	& -2 \sum_{\tilde{a}=0}^{1} p(\tilde{a}) \log p(\tilde{a}) \; .
\end{eqnarray}
Noting that $p(a) = 2 p(\tilde{a})$ for $a,\tilde{a} \in \{0,1\}$, we obtain
\begin{eqnarray}
H(A)	& =	& -\sum_{a=0}^{1} p(a) \log p(a) \nonumber\\
		& =	& -2 \sum_{\tilde{a}=0}^{1} p(\tilde{a}) \log \left[2 p(\tilde{a})\right] \nonumber\\
		& =	& -2 \sum_{\tilde{a}=0}^{1} p(\tilde{a}) -2 \sum_{\tilde{a}=0}^{1} p(\tilde{a}) \log p(\tilde{a}) \nonumber\\
\label{eq:ea}	& =	& -2 \sum_{\tilde{a}=0}^{1} p(\tilde{a}) +H(\widetilde{A}) \; .
\end{eqnarray}
Now let us expand the joint entropy of $\widetilde{A}$ and $E$ to ultimately show that the difference in the entropies of $\widetilde{A}$ and $A$ is exactly cancelled by the same difference in the joint entropies. The joint entropy is the average of joint entropies of Eve's different basis choices, $H_z(\widetilde{A},E)$ and $H_x(\widetilde{A},E)$, which are in fact equal. That is,
\begin{eqnarray}
H(\widetilde{A},E)	& =	& \frac{1}{2} H_z(\widetilde{A},E) + \frac{1}{2} H_x(\widetilde{A},E) \nonumber \\
							& =	& -\frac{1}{2} \sum_{e=0}^1 \sum_{\tilde{a}=0}^{-} p(e|\tilde{a},P_\mathrm{E}=z) p(\tilde{a}) \log [p(e|\tilde{a},P_\mathrm{E}=z) p(\tilde{a})] \nonumber \\
							&		& -\frac{1}{2} \sum_{e=0}^1 \sum_{\tilde{a}=0}^{-} p(e|\tilde{a},P_\mathrm{E}=x) p(\tilde{a}) \log [p(e|\tilde{a},P_\mathrm{E}=x) p(\tilde{a})] \nonumber
\end{eqnarray}
\begin{eqnarray}
							& =	& -\frac{1}{2} \sum_{e=0}^1 \sum_{\tilde{a}=0}^{1} p(\tilde{a}) \big\{ [p(e|\tilde{a},P_\mathrm{E}=z) + p(e|\tilde{a},P_\mathrm{E}=x)] \log p(\tilde{a}) \nonumber \\
							&		& + p(e|\tilde{a},P_\mathrm{E}=z) \log p(e|\tilde{a},P_\mathrm{E}=z) + p(e|\tilde{a},P_\mathrm{E}=x) \log p(e|\tilde{a},P_\mathrm{E}=x) \big\} \nonumber \\
							& 	& -\frac{1}{2} \sum_{e=0}^1 \sum_{\tilde{a}=+}^{-} p(\tilde{a}) \big\{ [p(e|\tilde{a},P_\mathrm{E}=z) + p(e|\tilde{a},P_\mathrm{E}=x)] \log p(\tilde{a}) \nonumber \\
							&		& + p(e|\tilde{a},P_\mathrm{E}=z) \log p(e|\tilde{a},P_\mathrm{E}=z) + p(e|\tilde{a},P_\mathrm{E}=x) \log p(e|\tilde{a},P_\mathrm{E}=x) \big\} \nonumber \\
							& =	& -\frac{1}{2} \sum_{e=0}^1 \sum_{\tilde{a}=0}^{1} p(\tilde{a}) \big\{ [p(e|\tilde{a},B_{\mathrm{c}}) + p(e|\tilde{a},B_{\mathrm{w}})] \log p(\tilde{a}) \nonumber \\
							&		& + p(e|\tilde{a},B_{\mathrm{c}}) \log p(e|\tilde{a},B_{\mathrm{c}}) + p(e|\tilde{a},B_{\mathrm{w}}) \log p(e|\tilde{a},B_{\mathrm{w}}) \big\} \nonumber \\
							& 	& -\frac{1}{2} \sum_{e=0}^1 \sum_{\tilde{a}=+}^{-} p(\tilde{a}) \big\{ [p(e|\tilde{a},B_{\mathrm{w}}) + p(e|\tilde{a},B_{\mathrm{c}})] \log p(\tilde{a}) \nonumber \\
							&		& + p(e|\tilde{a},B_{\mathrm{w}}) \log p(e|\tilde{a},B_{\mathrm{w}}) + p(e|\tilde{a},B_{\mathrm{c}}) \log p(e|\tilde{a},B_{\mathrm{c}}) \big\} \nonumber \\
							& =	& -\sum_{e=0}^1 \sum_{\tilde{a}=0}^{1} p(\tilde{a}) \big\{ [p(e|\tilde{a},B_{\mathrm{c}}) + p(e|\tilde{a},B_{\mathrm{w}})] \log p(\tilde{a}) \nonumber \\
\label{eq:jetildeae}	&		& + p(e|\tilde{a},B_{\mathrm{c}}) \log p(e|\tilde{a},B_{\mathrm{c}}) + p(e|\tilde{a},B_{\mathrm{w}}) \log p(e|\tilde{a},B_{\mathrm{w}}) \big\} \; ,
\end{eqnarray}
where we have used the fact that the $z$ basis is correct for $\tilde{a}=0,1$ and incorrect for $\tilde{a}=+,-$, and conversely for the $x$ basis. The last equality follows from the equality of the sums over $\tilde{a}$.

For the joint entropy of $A$ and $E$ we use again the average over the cases of correct and incorrect measurement basis:
\begin{eqnarray}
H(A,E)	& =	& \frac{1}{2} H_c(A,E) + \frac{1}{2} H_w(A,E) \nonumber \\
			& =	& -\frac{1}{2} \sum_{e=0}^1 \sum_{a=0}^1 p(e|a,B_{\mathrm{c}}) p(a) \log [p(e|a,B_{\mathrm{c}}) p(a)] \nonumber \\
			&		& -\frac{1}{2} \sum_{e=0}^1 \sum_{a=0}^1 p(e|a,B_{\mathrm{w}}) p(a) \log [p(e|a,B_{\mathrm{w}}) p(a)] \nonumber \\
			& =	& -\frac{1}{2} \sum_{e=0}^{1} \sum_{a=0}^{1} p(a) \big\{ [p(e|a,B_{\mathrm{c}}) + p(e|a,B_{\mathrm{w}})] \log p(a)\nonumber \\
			&		& +p(e|a,B_{\mathrm{c}}) \log p(e|a,B_{\mathrm{c}}) + p(e|a,B_{\mathrm{w}}) \log p(e|a,B_{\mathrm{w}}) \big\} \nonumber
\end{eqnarray}
\begin{eqnarray}
			& =	& -\sum_{e=0}^{1} \sum_{\tilde{a}=0}^{1} p(\tilde{a}) \big\{ [p(e|\tilde{a},B_{\mathrm{c}}) + p(e|\tilde{a},B_{\mathrm{w}})] \log (2 p(\tilde{a})) \nonumber \\
			&		& +p(e|\tilde{a},B_{\mathrm{c}}) \log p(e|\tilde{a},B_{\mathrm{c}}) + p(e|\tilde{a},B_{\mathrm{w}}) \log p(e|\tilde{a},B_{\mathrm{w}}) \big\} \nonumber \\
			& \stackrel{(\ref{eq:jetildeae})}{=}		& -\sum_{e=0}^{1} \sum_{\tilde{a}=0}^{1} p(\tilde{a}) [p(e|\tilde{a},B_{\mathrm{c}}) + p(e|\tilde{a},B_{\mathrm{w}})] + H(\widetilde{A},E) \nonumber \\
			& \stackrel{(\ref{eq:totalprob})}{=}	& -\sum_{e=0}^{1} \sum_{\tilde{a}=0}^{1} p(\tilde{a}) \cdot 2 p(e|\tilde{a}) + H(\widetilde{A},E) \nonumber \\
			& \stackrel{(\ref{eq:bayes})}{=}		& -2\sum_{\tilde{a}=0}^{1} \sum_{e=0}^{1} p(e) p(\tilde{a}|e) + H(\widetilde{A},E) \nonumber \\
\label{eq:jeae}	&  \stackrel{(\ref{eq:totalprob})}{=}	& -2\sum_{\tilde{a}=0}^{1} p(\tilde{a}) + H(\widetilde{A},E) \; .
\end{eqnarray}
Equations (\ref{eq:itildeae}), (\ref{eq:ea}), and (\ref{eq:jeae}) yield
\begin{eqnarray}
I(A,E)	& =	& H(E) + H(A) - H(A,E) \nonumber \\
			& =	& H(E) - 2 \sum_{\tilde{a}=0}^{1} p(\tilde{a}) +H(\widetilde{A}) + 2\sum_{\tilde{a}=0}^{1} p(\tilde{a}) - H(\widetilde{A},E) \nonumber \\
			& =	& H(E) +H(\widetilde{A}) - H(\widetilde{A},E) \nonumber \\
\label{eq:iaeeq}	& =	& I(\widetilde{A},E) \; ,
\end{eqnarray}
which completes our proof. Note that in obtaining Eq.~(\ref{eq:iaeeq}) we used only three assumptions:
\begin{itemize}
\item[i)] $p(\widetilde{A} =0) = p(\widetilde{A} = +)$ and $p(\widetilde{A} = 1) = p(\widetilde{A} = -)$.
\item[ii)] $p(a) = 2 p(\tilde{a})$ for $a, \tilde{a} \in \{0,1\}$.
\item[iii)] For $\tilde{a} = 0,1$, the $z$ basis is correct and the $x$ basis incorrect, and vice versa for $\tilde{a}=+,-$.
\end{itemize}

Next, we analyze how much errors Eve's strategy induces to Bob's sifted key, i.e., we calculate the QBER that Alice and Bob observe, given that Eve uses the IR strategy. The error rate is defined as the probability that an individual bit value in Bob's sifted key differs from the corresponding value in Alice's sifted key. Formally,
\begin{equation}
\label{eq:QBER}
\mathrm{QBER} = p(b \neq a) = \sum_a p(B=\bar{a}|A=a) p(a) \; ,
\end{equation}
where $a$ and $b$ are the outcomes of Alice and Bob's random variables $A$ and $B$, respectively. The bar over the symbol $\bar{a}$ represents an increment of 1 modulo 2, i.e., $\bar{0} = 1$ and $\bar{1} = 0$. The reader is reminded that Eve is in total control of the states that Bob receives, but has only partial control over Bob's measurement results, i.e., outcomes of $B$.

Because we are considering bits in the sifted key only, the correctness of Bob's result depends on Eve's basis choice:
\begin{eqnarray}
\mathrm{QBER}	& =												& \sum_a p(B=\bar{a}|A=a) p(a) \nonumber \\
							& \stackrel{(\ref{eq:probsum})}{=}	& \sum_a \underbrace{[p(B=\bar{a}, B_{\mathrm{c}}|A=a)}_{=0} + p(B=\bar{a}, B_{\mathrm{w}}|A=a)] p(a) \nonumber \\
							& \stackrel{(\ref{eq:cond1})}{=}		& \sum_a p(B=\bar{a}| B_{\mathrm{w}}, A=a) p(B_{\mathrm{w}}| A=a) p(a) \nonumber \\
							& =												& \frac{1}{2} \sum_a p(B=\bar{a}| B_{\mathrm{w}}, A=a) p(B_{\mathrm{w}}) \nonumber \\
							& =												& \frac{1}{4} \sum_{a=0}^{1} p(B=\bar{a} | B_{\mathrm{w}}, A=a) \nonumber \\
							& =												& \frac{1}{4} \Big(\frac{1}{2} + \frac{1}{2}\Big) = \frac{1}{4} \; ,
\end{eqnarray}
where we have used the fact that when Eve's basis choice is wrong, i.e., incompatible with Alice's choice, it is also incompatible with Bob's choice, in which case Bob gets an incorrect result with probability $\frac{1}{2}$.

In summary, we have shown that the IR attack strategy gives Eve 0.5 bits of information per interception and induces an average QBER of 25\% in the sifted key. In practice, a 25\% QBER would probably be considered too high by Alice and Bob, and they would thus abort the protocol. However, Eve does not have to intercept every transmission, instead, she can choose to interfere with only a fraction $0 \leq \xi \leq 1$ of the transmissions. Then Eve's information as well as the QBER is linearly parametrized by $\xi$:
\begin{eqnarray}
I_{A,E}(\xi)				& =	& \xi/2 \; , \\
\mathrm{QBER}(\xi)	& =	& \xi/4 \; ,
\end{eqnarray}
from which the maximal information Eve can gain for a given QBER $q$ is
\begin{equation}
%\begin{displaymath}
I_{max}^{\mathrm{IR}}(q) = \left\{
\begin{array}{ll}
2q			& \textrm{if $0\leq q \leq 1/4$}\\
0.5 \quad	& \textrm{if $1/4 < q \leq 1/2$} \; .
\end{array}
\right.
%\end{displaymath}
\end{equation}

\subsubsection{Incoherent attack}
In an incoherent or individual attack, Eve entangles each transmitted qubit individually to a \emph{probe}. Eve's probes are quantum systems capable of retaining their state until the basis reconciliation phase. Alternatively, the state of each probe can be kept in separate quantum memory\footnote{Long-term quantum memory is a delicate issue in its own right. Photonic physical-qubit memories are discussed, for example, in Refs.~\cite{quantmem1, quantmem2}.}. The probes are assumed to be identical, and there is one probe for each eavesdropped transmission. Four-dimensional, i.e., two-qubit, probes are sufficient for Eve's purposes in BB84 \cite{pra53fuchs}. After basis reconciliation, Eve measures the probe states one-by-one in an attempt to gain as much information as possible on the sifted key. The qubit-probe interaction $\mathcal{U}$ can be assumed unitary\footnote{A non-unitary interaction would be equivalent to a unitary one, only with a higher-dimensional probe.} and independent of the state of the qubit. The interaction can be viewed as an act of transferring information from the transmitted qubit to one or more probe qubits. For Eve, the optimal choices of $\mathcal{U}$ are parametrized by a real variable $\eta$. Therefore, variable $\eta$ actually parametrizes a whole---uncountably infinite---family of attacks referred to as incoherent attacks.

The maximal mutual information that Eve can gain with an incoherent attack is
\begin{eqnarray}
I_{max}^{\mathrm{incoh.}}(q)	& =	& \left(\frac{1}{2} + \sqrt{q(1-q)} \right) \log\big(1+2\sqrt{q(1-q)}\big) \nonumber\\
											&		& + \left(\frac{1}{2} - \sqrt{q(1-q)} \right) \log\big(1-2\sqrt{q(1-q)}\big) \;, \label{eq:incohmax}
\end{eqnarray}
where is $q$ is a given QBER. It is also known that an interaction $\mathcal{U}$ and a probe measurement scheme achieving this bound exist \cite{pra56fuchs}.

An incoherent attack achieving the bound in Eq.~(\ref{eq:incohmax}) is equivalent to optimal cloning of the transmitted qubit. The optimal cloner is the $1 \to 2$ phase-covariant qubit cloner defined by transformation (\ref{eq:phcovcloner}), which is justified as follows. We only consider transmissions that contribute to the sifted key. We assume that Eve attempts to clone state $|+\rangle$, since the calculation is similar for all other BB84 states. Setting $\varphi=0$, Eq.~(\ref{eq:phcovtrans}) yields the result of the cloning process as
\begin{equation}
|C\rangle = \mathcal{U}|+\rangle|0\rangle = \frac{1}{\sqrt{2}}\big(|0\rangle|0\rangle + \cos(\eta) |1\rangle|0\rangle + \sin(\eta) |0\rangle|1\rangle \big) \; .
\end{equation}
Eve then sends Bob the qubit in the first slot and keeps the qubit in the second slot for herself. When Bob measures his qubit, he gets the correct result, i.e., zero in the $x$ basis, with probability
\begin{equation}
\label{eq:incohcorrprob}
p(B=0) = \langle C| \big(P_0^x\otimes I\big) |C\rangle = \frac{1}{2}\big(1+\cos\eta\big) \; .
\end{equation}
Hence, the QBER is
\begin{equation}
\label{eq:incohqber}
q(\eta) = \frac{1}{2}\big(1 - \cos\eta\big) \; .
\end{equation}

The best strategy for Eve is to simply measure her probe qubit the same way she measures transmitted qubits in the IR attack. However, because Eve can keep her probes intact until she learns which basis Alice has used in the transmission, she knows in which basis to perform the measurement. Eve gets the correct result in the $x$ basis with probability
\begin{equation}
\label{eq:incoheveprob}
P_{E=0}(\eta) = \frac{1}{2}\big(1+\sin\eta\big) \; .
\end{equation}
Eve's mutual information on the key decreases with increasing uncertainty in her measurement result. That is,
\begin{equation}
\label{eq:incohinfo}
I_{A,E}(\eta) = 1 - H_{\mathrm{bin}}[P_{E=0}(\eta)] \; .
\end{equation}
Eliminating $\eta$ in Eqs.~(\ref{eq:incohqber}) and (\ref{eq:incohinfo}) yields the bound in Eq.~(\ref{eq:incohmax}). Note the equivalence of Eqs.~(\ref{eq:incohcorrprob}), (\ref{eq:incoheveprob}) and the fidelities given in Eq.~(\ref{eq:phcovfids}). For a more detailed description of the cloning process, see Ref.~\cite{scarani}.

\begin{figure}[hbt]
\vspace{5mm}
\begin{center}
\includegraphics[width=0.6\textwidth]{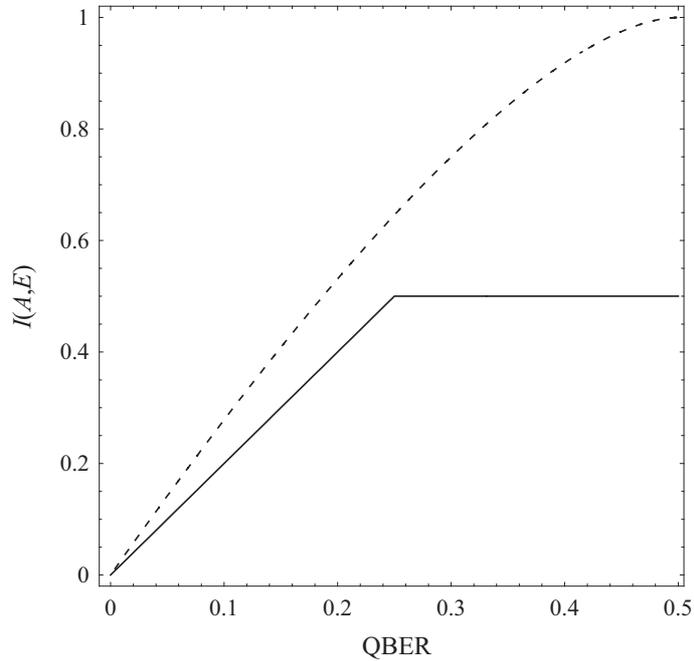}
\caption{Eve's maximal information on Alice's sifted key for a given QBER for intercept-resend (solid line) and incoherent (dashed line) attack strategies against the BB84 protocol.}
\label{fig:bb84miqber}
\end{center}
\end{figure}

Interacting with only a fraction $\xi$ of the transmissions does not provide Eve any advantage. This is because the mutual-information bound in Eq.~(\ref{eq:incohmax}) is a concave function of $q$, and hence, for a fixed QBER, adjusting the parameter $\eta$ and probing every transmission is always at least as beneficial as not probing every transmission. Figure \ref{fig:bb84miqber} shows the maximal mutual information as a function of a given QBER for the incoherent and intercept-resend attacks.

%equivalence to cloning attack
%\small{T\"ah\"an ainakin v\"ah\"an. Ehka enemman riippuen siit\"a ehdit\"a\"anko viela saada vastaavalle hyokkaykselle meidan protokollassa jotain tuloksia.}

%\small{Graafi jossa plotattu Even mahdollisuudet eri hyokkayksilla: Even Mutual information vs. QBER.}

\subsubsection{Coherent attack}
In a coherent attack, Eve is in possession of an unlimited-dimensional probe in an arbitrary initial state. Eve is allowed to apply any unitary transformation to the entire transmitted qubit sequence and the probe. The probe state is retained until all public discussions between Alice and Bob are finished, and Eve is then allowed to perform arbitrary measurements on the probe system as a whole. Collective attacks are a subclass of coherent attacks, in which Eve is allowed to entangle the qubits and probes individually but still use any conceivable measurement scheme after Alice and Bob's public discussions. \cite{bouwmeester}

For coherent attacks, the various security proofs state that the probability, that Alice and Bob unknowingly agree on a key that Eve has more than an exponentially small amount of information, is exponentially small in some security parameter under Alice and Bob's control. No explicit maximal mutual information for a given QBER has been presented in the literature. \cite{sciencelo, secproof1, secproof2, secproof3, secproof4, secproof5, secproof6}

%\small{T\"ast\"a aika v\"ah\"an joka tapauksessa.}

%gravitational attacks?

\subsection{Privacy amplification}
Because an error-free quantum channel does not exist, Alice and Bob have to work with some finite QBER. Consequently, they can never be absolutely certain that Eve has not eavesdropped parts of the generated key. Given a long enough key sequence, however, Alice and Bob can shorten the key and reduce Eve's information on it to an arbitrarily low value by public classical communication. This procedure is called privacy amplification.

The essential step in privacy amplification algorithms is typically the following: Alice randomly chooses a pair of slots $\{i,j\}$ in the error-corrected sifted key and informs Bob of her choice. Both participants then calculate $c_{ij} = \mathrm{XOR}(a_i,a_j)$. Alice and Bob obtain the same value for $c_{ij}$, since their bit strings are identical. They then replace the bits in slots $i$ and $j$ with the value $c_{ij}$. Any uncertainty Eve has about the bit values in the slots is always increased by this process. For example, if Eve only knows the value of the bit in slot $i$, after privacy amplification she knows nothing of the value of slot $i$. This step is iterated for as long as is necessary to bring Eve's maximal information on the key to a low enough value. More sophisticated protocols work on larger bit blocks. \cite{gisin}

%\small{ Maininta ja lyhyt esimerkki privacy amplificationista.}
%privacy amplification

\section{Other protocols}
\label{sec:qkdother}

\subsection{Einstein-Podolsky-Rosen protocol}
In 1991, A.\ Ekert published the Einstein-Podolsky-Rosen (EPR) QKD protocol, sometimes referred to as E91 \cite{modEPR}. This protocol is named after the famous EPR thought experiment constructed to prove that quantum mechanics is not a complete description of reality \cite{epr,nielsenchuang}. In the EPR protocol, Alice and Bob do not send quantum states to each other, but instead rely on a third party, Charlie, to transmit two qubits in an entangled state, one qubit to Alice and the other to Bob. Specifically, the state that Charlie emits is
\begin{equation}
\label{eq:eprcharlie}
|\Phi^+\rangle = \frac{1}{\sqrt{2}} (|0\rangle|0\rangle + |1\rangle|1\rangle) \; ,
\end{equation}
also known as one of the Bell states. A pair of qubits in this state are said to form an \emph{EPR pair}. Upon reception of the qubits, Alice and Bob randomly choose between two measurement bases, just as in BB84. Later the bases are announced in public, and the sifted key is obtained by discarding the results for which the bases did not match.

Alice and Bob can perform a test to see whether Charlie truly emits the state in Eq.~(\ref{eq:eprcharlie}). This test is based on Bell's inequality which demonstrates that a local theory cannot give the correlations that quantum mechanics predicts \cite{bell}. To be completely assured that Eve has not tampered with the emitted state, Alice and Bob must observe a maximal violation of Bell's inequality. In practice, because of noise or eavesdropping, only a sub-maximal violation is observed, requiring the use of error correction and privacy amplification for obtaining a secret key. When Charlie emits state $|\Phi^+\rangle$, this protocol is equivalent to BB84 \cite{gisin}.
%Intuitively, in BB84, it is as if the qubit from Charlie to Alice first propagates backwards in time, and then forwards in time from Charlie to Bob

\subsection{Two-state protocol}
In 1992, C. H. Bennett proposed a simple variant of the original BB84 protocol \cite{mod2state}. It is known as B92 or the two-state protocol. The latter name comes from the essential modification to BB84. The four states $\{|0\rangle, |1\rangle, |+\rangle, |-\rangle\}$ used in BB84 are more than is necessary for Eve not being able to eavesdrop the transmissions without being noticed. In fact, using only two non-orthogonal states suffices, e.g., $|0\rangle$ and $|+\rangle$. In the B92 protocol, Alice randomly chooses which one of the two states she transmits, and Bob randomly chooses a measurement basis for each reception. The rest of the protocol is identical to BB84. The requirement of transmitting only two different states renders the experimental implementation of the protocol less demanding. Although Eve still inevitably perturbs the transmission if she interferes with it, she can unambiguously distinguish between the two states at the cost of some transmissions being lost completely. \cite{gisin}

\subsection{Six-state protocol}
Another fairly simple variant of BB84 is the six-state protocol proposed by D. Bru\ss{} in 1998 \cite{mod6state}. The six-state protocol uses three conjugate bases for the quantum channel transmissions: not only the eigenbases of Pauli matrices $\sigma_x$ and $\sigma_z$ but also the eigenbase of $\sigma_y$. Alice randomly transmits and Bob randomly measures in one of these bases. The intercept-resend strategy induces a 33\% QBER in this protocol. If Eve employs an incoherent attack against this protocol, then given a QBER $q$, her maximal information on the key is
\begin{eqnarray}
I_{max}^{\mathrm{incoh.}}(q)	& =	& 1 - (1-q) H_{\mathrm{bin}}[g(q)] \; , \; \mathrm{where} \\
&& \nonumber \\
g(q)											& =	& \frac{1}{2} \left(1 + \frac{1}{1-q} \sqrt{q(2 - 3q)} \right) \; .
\end{eqnarray}
This is less than, although close to, the maximum in Eq.~(\ref{eq:incohmax}) for $0 < q < 0.5$. That is, to achieve the same information on the key, Eve must induce a slightly higher QBER than in the original BB84. However, because Alice and Bob use three different bases, Bob chooses the correct basis on average only $\frac{1}{3}$ of the time. Therefore, to generate a sifted key of given length, more quantum transmissions are needed than in the BB84 protocol.

\subsection{Adjusted basis probabilities protocol}
In the original BB84 protocol, the two transmission and measurement bases $z$ and $x$ are chosen with equal probabilities. In 1998, M.\ Ardehali \emph{et al.} proposed a variant in which the use of one of the bases has a significantly higher probability \cite{modbasesprob}. The adjusted probability is announced in public. The advantage of this modification is that a considerably smaller amount of measurement results need to be discarded in the basis reconciliation phase. However, Eve's information on the key is higher, since she can always employ the basis that is used more frequently. To counteract this, the authors suggest a sophisticated error analysis scheme. It is not clear whether this modification ultimately improves on the efficiency of BB84.

%\small{Olennaiset ominaisuudet n\"aist\"a: BB84: (2-state, 6-state, eri todenn\"akoisyydet eri l\"ahetyskannoilla), EPR, not(SARG)}

%\small{Edell\"amainituista graafi: Even Mutual information vs. QBER.} ei

%% file: analysis.tex
\chapter{Analysis and Results}
\label{ch:analysis}

This chapter describes in detail our proposed amendment to the BB84 protocol. The purpose of the modification is to yield Alice and Bob advantage against an eavesdropper in terms of mutual information. As a demonstration of the idea behind the modification, we present an analysis of the difficulty of approximating an entangled state of two qubits with two product-state qubits. As our main result, we give explicit bounds on the information of an eavesdropper employing an intercept-resend attack against our protocol as a function of the qubit error rate. We also discuss a special kind of attack against this protocol, one in which Eve recreates destroyed entanglement using EPR pairs.
%In addition, we compare the gained advantage to other variations of the BB84 protocol, and discuss the applicability and implementation of our proposed modification.

\section{Proposed modification to the BB84 protocol}
\label{sec:propmod}

We analyze a protocol based on the BB84 protocol. Our protocol differs from the original one in the following:
\begin{enumerate}
\item Prior to the key distribution, Alice and Bob publicly agree on an $N$-qubit unitary transformation $U\!:\mathcal{H}^N \to \mathcal{H}^N$.
\item Alice's actions differ from BB84 such that she
\label{item:alicediffer}
	\begin{enumerate}
	\item \label{item:apost1} postpones her transmissions\footnote{Postponing the processing of existing qubits in items \ref{item:alicediffer}(a), \ref{item:alicediffer}(c), and \ref{item:bobdiffer}(a) requires that Alice and Bob employ short-term quantum memory.} until she has generated $N$ qubits to transmit,
	\item then applies $U$ to the qubits, and
	\item \label{item:apost2} transmits them one at a time, always waiting for Bob to acknowledge the reception of the previous qubit before sending the next one.
	\end{enumerate}
\item Bob's actions differ from BB84 such that he
\label{item:bobdiffer}
	\begin{enumerate}
	\item \label{item:bpost} postpones his measurements until $N$ qubits have arrived,
	\item immediately acknowledges each received qubit to Alice, and
	\item having received a sequence of $N$ qubits, applies $U^{-1} = U^{\dagger}$ to the qubits and measures them exactly as in BB84.
	\end{enumerate}
\end{enumerate}

The transformation $U$ can be viewed as an extension or a plug-in to the BB84 protocol. Without Eve's interference, the use of $U$ and $U^{-1}$ is fully transparent from Alice's and Bob's point of view. Note that Eve is fully aware of $U$, since it is announced in public. Because Bob acknowledges every arrived qubit, Eve has only one-by-one access to the particles of the $N$-qubit state. The transformation is
\begin{equation}
U (|a_1\rangle\otimes|a_2\rangle\otimes\cdots\otimes|a_N\rangle) = |\psi_{a_1,a_2,\ldots,a_N}\rangle \; ,
\end{equation}
where $|a_i\rangle$ are the states of the original BB84 protocol, i.e., $|a_i\rangle \in \{|0\rangle, |1\rangle, |+\rangle, |-\rangle\}$. In our modified protocol, Alice sends the qubits of the state $|\psi_{a_1,a_2,\ldots,a_N}\rangle$ to Bob one at a time.

If $U$ is of the form
\begin{equation}
\label{eq:udekompo}
U = U_1\otimes U_2\otimes \cdots\otimes U_N \; ,
\end{equation}
where $U_j$ are single-qubit gates, i.e, $U_j\!:\mathbb{C}^2 \to \mathbb{C}^2$ for $j \in \{1,2,\ldots,N\}$, then
\begin{eqnarray}
U (|a_1\rangle\otimes|a_2\rangle\otimes\cdots\otimes|a_N\rangle)	& =	& (U_1\otimes U_2\otimes \cdots\otimes U_N) (|a_1\rangle\otimes|a_2\rangle\otimes\cdots\otimes|a_N\rangle) \nonumber \\
																									& =	& U_1|a_1\rangle \otimes U_2 |a_2\rangle \otimes \cdots \otimes U_N |a_N\rangle \nonumber \\
																									& =	& |\psi_{a_1}\rangle \otimes |\psi_{a_2}\rangle \otimes \cdots \otimes |\psi_{a_N}\rangle \label{eq:uprodstate} \; .
\end{eqnarray}
In other words, if $U$ is decomposable to single-qubit gates, the transmitted $N$-qubit state is a product state. Given the product state, Eve can perfectly undo $U$, attack the individual unentangled qubits, and reconstruct the transmitted state by using the single-qubit gates $U_j^{\dagger}$ and $U_j$. Therefore, Alice and Bob should choose $U$ such that it produces an entangled $N$-qubit state. This implies
\begin{equation}
U \neq U_1\otimes U_2\otimes \cdots\otimes U_N \; .
\end{equation}
That is, by using a non-local $U$, Alice and Bob utilize entanglement to prohibit Eve from fully accessing the transmitted qubits. In the following sections we restrict our analysis to the case $N=2$.

%\small{Tarkka ja formaali selitys meidan protokollasta.}
%\small{Vahan siita miten entanglementti on se oleellinen juttu tassa tyossa.}

\section{Product-state approximation of an entangled qubit pair}

To demonstrate the underlying idea in using an entangling $N$-qubit gate in BB84, we perform an analysis on how closely an entangled two-qubit state can be approximated with two product-state qubits. This analysis shows that even if perfect cloning of quantum states was possible, the protocol poses an inherent limitation for Eve.

Assume that Eve constructs the state
\begin{equation}
\bigotimes_{i=1}^N |\psi_i\rangle \; , \quad |\psi_i\rangle \in \mathbb{C}^2 = \mathcal{H}^1 \; , \left|\left||\psi_i\rangle\right|\right|=1 \; ,
\end{equation}
in an attempt to approximate a normalized state $|\psi\rangle \in \mathbb{C}^{2^N} = \mathcal{H}^N$ being transmitted one qubit at a time from Alice to Bob. Eve tries to minimize the error in this approximation whereas
%\begin{equation}
%E_{min} = \min_{\{|\psi_i\rangle\}} \big|\big||\psi\rangle - \bigotimes_{i=1}^N |\psi_i\rangle \big|\big| \; .
%\end{equation}
Alice and Bob want to maximize Eve's minimal error by choosing $|\psi\rangle$ appropriately. This maximal-minimal error is
\begin{equation}
\label{eq:emaxmin}
E_{\mathrm{mm}} := \max_{|\psi\rangle} \min_{\{|\psi_i\rangle\}} \Big|\Big||\psi\rangle - \bigotimes_{i=1}^N |\psi_i\rangle \Big|\Big| \; .
\end{equation}
%appropriately wiktionary

\subsection{Theory}
We assume that Alice and Bob have chosen $N=2$. We write the state Alice uses as
\begin{equation}
|\psi\rangle =	\left( \begin{array}{cccc} r_{\alpha_1}e^{i\alpha_1}	& r_{\alpha_2}e^{i\alpha_2} & r_{\alpha_3}e^{i\alpha_3} & r_{\alpha_4}e^{i\alpha_4} \end{array} \right)^{\mathrm{T}} \; ,
\end{equation}
and the states Eve uses as
\begin{eqnarray}
|\psi_1\rangle	& =	& \left( \begin{array}{cc} r_{\phi_1}e^{i\phi_1}				& r_{\phi_2}e^{i\phi_2} \end{array} \right)^{\mathrm{T}} \; , \\
|\psi_2\rangle	& =	& \left( \begin{array}{cc} r_{\omega_1}e^{i\omega_1}	& r_{\omega_2}e^{i\omega_2} \end{array} \right)^{\mathrm{T}} \; .
\end{eqnarray}
Normalization of the state vectors implies
\begin{eqnarray}
r_{\alpha_1}^2 + r_{\alpha_2}^2 + r_{\alpha_3}^2 + r_{\alpha_4}^2		& =	& 1 \; , \\
r_{\phi_1}^2 + r_{\phi_2}^2																& =	& 1 \; , \\
r_{\omega_1}^2 + r_{\omega_2}^2													& =	& 1 \; .
\end{eqnarray}
The moduli $r_{\alpha_j}$ are conveniently parametrized by three angles $\bar{\theta} = (\theta_1,\theta_2,\theta_3)$ as the surface of a four-dimensional sphere:
\begin{equation}
\left \{
\begin{array}{lcl}
r_{\alpha_1}	& =	& \cos \theta_1 \\
r_{\alpha_2}	& =	& \sin \theta_1 \cos \theta_2  \\
r_{\alpha_3}	& =	& \sin \theta_1 \sin \theta_2 \cos \theta_3 \\
r_{\alpha_4}	& =	& \sin \theta_1 \sin \theta_2 \sin \theta_3 \; .
\end{array}
\right .
\end{equation}
The moduli of Eve's qubits represent two circles for which two angles $\Phi$ and $\Omega$ suffice as
\begin{eqnarray}
&& r_{\phi_1} = \cos \Phi \quad \mathrm{and} \quad r_{\phi_2} = \sin \Phi \; , \\
&& r_{\omega_1} = \cos \Omega \quad \mathrm{and} \quad r_{\omega_2} = \sin \Omega \; .
\end{eqnarray}

After several simplifying steps, one obtains
\begin{equation}
E_{\mathrm{mm}} = \big\{ 2[1 - \min_{\bar{\theta}, \bar{\alpha}} \max_{\Phi, \Omega, \bar{\phi}, \bar{\omega}} G(\bar{\theta}, \bar{\alpha}, \Phi, \Omega, \bar{\phi}, \bar{\omega})] \big\}^{1/2} \; ,
\end{equation}
where the complex argument parameters are gathered into vectors
\begin{equation}
\bar{\alpha} = (\alpha_1, \alpha_2, \alpha_3, \alpha_4) \; ; \quad \bar{\phi} = (\phi_1, \phi_2) \; ; \quad \bar{\omega} = (\omega_1, \omega_2) \;.
\end{equation}
The function $G$, which Eve tries to maximize and whose maximum Alice and Bob attempt to minimize, is
\begin{eqnarray}
G(\bar{\theta}, \bar{\alpha}, \Phi, \Omega, \bar{\phi}, \bar{\omega})& :=	& \cos \Phi \,[ \cos \Omega \cos(\alpha_1 - \phi_1 - \omega_1) \cos \theta_1 \nonumber\\
&&																									+ \sin \Omega \cos(\alpha_2 - \phi_1 - \omega_2) \sin \theta_1 \cos \theta_2] \nonumber\\
&&																									+ \sin \Phi \,[ \cos \Omega \cos(\alpha_3 - \phi_2 - \omega_1) \sin \theta_1 \sin \theta_2 \cos \theta_3 \nonumber\\
&&																									+ \sin \Omega \cos(\alpha_4 - \phi_2 - \omega_2) \sin \theta_1 \sin \theta_2 \sin \theta_3 ] \; .
\end{eqnarray}
Global bounds for the error follow from the extreme values of $G$
\begin{equation}
-1 \leq G \leq 1 \quad \Rightarrow \quad 0 \leq E_{\mathrm{mm}} \leq 2 \; .
\end{equation}

Because Alice and Bob wish to maximize the norm in Eq.~(\ref{eq:emaxmin}), it is of no use to consider parameters in the set $\left\{\bar{\theta}, \bar{\alpha}\right\}$ that have no effect on the minimal value Eve is trying to achieve. We show that it is in fact sufficient to consider the maximization with three of the phases $\alpha_j$ fixed---varying them cannot increase the minimum. Firstly, the global phase of the pair $|\psi\rangle$ offers Alice and Bob no advantage, as it is directly reproduced by Eve. Secondly, Eve can apply any single-qubit gates to $|\psi_1\rangle$ and $|\psi_2\rangle$. For instance, Eve can freely choose $\beta_1,\beta_2,\beta_3 \in \mathbb{R}$ and apply the gate
\begin{equation}
e^{i \beta_0} \left( e^{i\beta_1\sigma_z} \otimes e^{i\beta_2\sigma_z} \right) = \left(\begin{array}{cccc}
	e^{i(\beta_0 + \beta_1 + \beta_2)}	& 0	& 0	& 0 \\
	0	& e^{i(\beta_0 + \beta_1 - \beta_2)}		& 0	& 0 \\
	0	& 0	& e^{i(\beta_0 - \beta_1 + \beta_2)}		& 0 \\
	0	& 0	& 0	& e^{i(\beta_0 - \beta_1 - \beta_2)} \\
\end{array} \right) \; 
\end{equation}
to her qubit pair. The global phase shift is implemented by $\beta_0$. By choosing the $\beta_j$ as
\begin{equation}
\left\{ \begin{array}{lcl}
	\beta_0 + \beta_1 + \beta_2	& =	& \alpha_1 \\
	\beta_0 + \beta_1 - \beta_2	& =	& \alpha_2 \\
	\beta_0 - \beta_1 + \beta_2	& =	& \alpha_3 \\
\end{array} \right.
\quad \Leftrightarrow \quad
\left\{ \begin{array}{lcl}
	\beta_0	& =	& \frac{1}{2}\left(\alpha_2 + \alpha_3\right) \\
	\beta_1	& =	& \frac{1}{2}\left(\alpha_1 - \alpha_3\right) \\
	\beta_2	& =	& \frac{1}{2}\left(\alpha_1 - \alpha_2\right)
\end{array} \right. \; ,
\end{equation}
Eve can reproduce the phases $\alpha_1, \alpha_2, \alpha_3$ in $|\psi\rangle$. Therefore, Alice and Bob may as well fix their value. Similar reasoning can be applied to the amplitudes $r_\Delta$, which is, however, not carried out here.

\subsection{Solution}

Having fixed $\alpha_1 = \alpha_2 = \alpha_3 = 0$, we solve for $E_{\mathrm{mm}}$, optimizing over $\bar{\theta}, \alpha_4$ and $\Phi, \Omega, \bar{\phi}, \bar{\omega}$. We use the built-in numerical optimization function of \textit{Mathematica} version 5.1.0.0 by Wolfram Research, Inc. As with any numerical maximization or minimization method, there are no guarantees that the found optimum is the global optimum. As discussed below, however, it is likely that a global optimum is found.

For the maximization of $G$ over Eve's parameters, we employ the \textit{RandomSearch} optimization method which generates, in this case, 100 random parameter starting points for the standard \textit{FindMinimum} function. \textit{RandomSearch} is a suitable method for maximizing $G$, since $G$ is a continuous and smooth function in all its arguments \cite{mathrs}. The minimization of this maximum is performed with the \textit{SimulatedAnnealing} method. Simulated annealing is a well-known optimization method that has similarities to the process of a physical system cooling down. First, the method randomly generates a set of starting points for the parameters. It also generates a random direction in the parameter space for each point. If moving to the selected direction satisfies the optimization goal better, the move is accepted, whereas if the move satisfies the goal worse, it is accepted with probability $p_m$. The probability $p_m$ decreases with each iteration, and also depends on how well the move satisfies the optimization goal. This procedure is repeated until the method stays at the same point for sufficiently many iterations, or until a predefined number of iterations is exceeded. Simulated annealing is a universally valid optimization method.

\subsection{Results}

We find that $E_{\mathrm{mm}} = 0.673$. This is achieved by choosing $\bar{\theta} = (1.228, 0.848, -0.499)$ and $\alpha_4 = 0.474$ with $\bar{\alpha} = (0,0,0,\alpha_4)$. To further confirm the result, the maximization of $G$ over Eve's parameters was also performed with the differential evolution, Nelder-Mead, and simulated annealing methods, in addition to \textit{RandomSearch}. The details of the methods are not discussed here, for more information, see, e.g., Ref.~\cite{mathrs}. Many different initial values were also tested. All these methods and all tested initial values resulted in the same maximum for $G$. Therefore, we are confident that we indeed have obtained the global maximum of $G$ for the values of $\bar{\theta}$ and $\bar{\alpha}$ given above. The optimal choice of Eve's parameters is not unique. Due to finite computing resources, no such intensive testing was applied to the more demanding minimization of the maximum of $G$ over $\bar{\theta}$ and $\bar{\alpha}$. Hence, we only state that the obtained values for $\bar{\theta}$ and $\bar{\alpha}$ are the optimal choice for Alice and Bob with high probability. That is, we settle for the confidence the simulated annealing method provides.

The obtained optimal values for $\bar{\theta}$ and $\bar{\alpha}$ approximately correspond to the state
\begin{equation}
\label{eq:optpsi}
|\psi\rangle = \left( \begin{array}{cccc} 0.34	& 0.62	& 0.62 & -0.30 - 0.15i \end{array} \right)^{\mathrm{T}} \; .
\end{equation}
The best Eve can do to approximate $|\psi\rangle$ with her two unentangled qubits is to choose, for example, $\Phi = 2.365$, $\Omega = 0.797$, $\phi_1 = 1.243$, $\phi_2 = 3.034$, $\omega_1 = 2.801$, and $\omega_2 = 1.472$. This approximately corresponds to the state
\begin{equation}
|\psi_1\rangle \otimes |\psi_2\rangle = \left( \begin{array}{cccc} 0.31 + 0.39i	& 0.46 - 0.21i	& 0.44 - 0.21i	& -0.10 - 0.49i
																	\end{array} \right)^{\mathrm{T}} \; .
\end{equation}
To recapitulate, if Alice and Bob provide Eve with the state in Eq.~(\ref{eq:optpsi}), it is guaranteed that the error in Eve's approximation is at least $0.673$.
%&& = \left( \begin{array}{cc} -0.23 - 0.68i	& -0.70 - 0.08i \end{array} \right)^{\mathrm{T}} \otimes \left( \begin{array}{cc} -0.66 + 0.23i	& 0.07 + 0.71i \end{array} \right)^{\mathrm{T}} \nonumber \\

%\small{Siita mista puhuttiin nyt viimeksi eli miten kahdella product state -qubitilla ei paasta kuin niin ja niin lahelle mielivaltaista 2-qubittitilaa (ja Eve minimoi etaisyytta ja Alice ja Bob maksimoi).}

%-----------------------------------------------------------------------------------------------------------------------------------------------------------------------

\section{Analysis of an intercept-resend attack}
In this section, we aim at answering the question: ``Assuming Eve uses the intercept-resend attack strategy, which $U$ should Alice and Bob choose?'' In the BB84 protocol, the IR attack strategy is less efficient than an optimal incoherent attack. However, it is not clear that the same holds for our augmented protocol. Moreover, if a transformation $U$ provides Alice and Bob advantage against an IR attack, it is likely that the advantage stands, at least to some extent, against more sophisticated attacks, as well.
%Therefore, it is well worthwhile to study how much advantage using a transformation $U$ gives Alice and Bob against an IR eavesdropper.

\subsection{Parametrization of $U$}
An arbitrary two-qubit gate has 16 degrees of freedom. For Alice and Bob's purposes however, several of these are useless. Firstly, one degree of freedom arises from the global phase shift introduced by the gate. It is well known that the global phase of the qubit pair is irrelevant. We can always choose the global phase such that the determinant of the gate is $+1$. Thus we can restrict our search to the special unitary group SU(4), the members of which have $4^2 -1 = 15$ degrees of freedom.

Following the treatment of J.\ Zhang \emph{et al.} \cite{pra67zhang}, we partition the group SU(4) into two: the subset of local gates, $\mathrm{L}_4 := \mathrm{SU(2)}\otimes\mathrm{SU(2)}$, and the subset of non-local gates, $\mathrm{NL}_4 := \mathrm{SU(4)}\backslash\mathrm{SU(2)}\otimes\mathrm{SU(2)}$. It is shown in Ref.~\cite{pra67zhang} that any $U \in \mathrm{SU(4)}$ can be decomposed as
\begin{eqnarray}
U	&=& k_2 A(c_1,c_2,c_3) k_1 \label{eq:su4decomp} \\
	&=& (k_{2,1} \otimes k_{2,2}) \, \exp \left[ \frac{i}{2} \left(c_1\, \sigma_x \otimes \sigma_x + c_2\, \sigma_y \otimes \sigma_y + c_3\, \sigma_z \otimes \sigma_z\right) \right] \, (k_{1,1} \otimes k_{1,2}) \; , \nonumber
\end{eqnarray}
where $k_i \in \mathrm{L}_4$ and thus $k_{i,j} \in \mathrm{SU(2)}$ and the parameters $c_l \in [0,\pi]\,,\; l = 1,2,3$.

%Let us represent the decomposition in Eq.~(\ref{eq:su4decomp}) as a quantum circuit.
Quantum circuits are a graphical way of representing quantum information processing, such as the application of a gate $k_2 A(c_1,c_2,c_3) k_1$ on two qubits. In a quantum circuit diagram, a single horizontal line represents a qubit. A double horizontal line represents a cbit. Time progresses from left to right, and an operation $O$ targeted to one or more qubits is shown as a box placed on top of the qubits involved in the operation $O$.
%Let a semicircle open to the left denote a projective measurement.

\begin{figure}[hbt]
\vspace{5mm}
\begin{center}
\includegraphics[height=23mm]{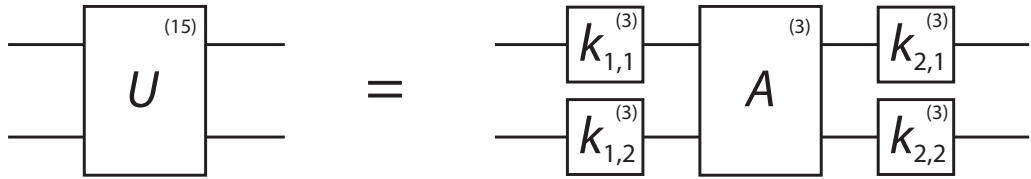}
\caption{Any $U \in \mathrm{SU(4)}$, shown left, is equal to a decomposition shown on the right. The number of degrees of freedom is shown in parentheses for each gate. }
\label{fig:su4decomp}
\end{center}
\end{figure}

\begin{figure}[hbt]
\vspace{5mm}
\begin{center}
\includegraphics[height=23mm]{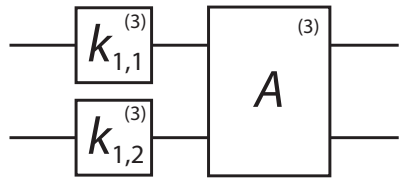}
\caption{A simplified gate model that is at least as useful for Alice and Bob as any other two-qubit transformation. The number of degrees of freedom is displayed in parentheses for each gate.}
\label{fig:su4useful}
\end{center}
\end{figure}

Figure \ref{fig:su4decomp} shows on the right the quantum circuit of the decomposed gate $U$. The qubits $|a_1\rangle, |a_2\rangle \in \{|0\rangle, |1\rangle, |+\rangle, |-\rangle\}$ generated by Alice enter the circuit from the left and depart at the right end after the application of $U$. Since we are interested in the security of this protocol, we assume that after the gate $U$ both qubits travel to Eve. Note that we do not assume that Eve necessarily does anything to either qubit. If Alice employs $k_2 = k_{2,1} \otimes k_{2,2}$, Eve can always undo and redo it perfectly with the single-qubit gates $k_{2,1}^{\dagger}$, $k_{2,2}^{\dagger}$ and $k_{2,1}$, $k_{2,2}$. Hence, $k_2$ is useless to Alice and Bob, and we may further restrict our search for a good gate $U$ to gates of the form $A(c_1, c_2, c_3) k_1$, shown in Fig.~\ref{fig:su4useful}. Thus we are left with 9 degrees of freedom for $U$.

%tähän ehkä vielä 1-qubittiportin parametrisointi

%\small{Selvennys miten mielivaltainen 2-qubittiportti dekomposoitiin 1-qubittiporteiksi ja \\yhdeksi 3:n vapausasteen 2-qubittiportiksi (J. Zhang phys rev A 67, 042313).}

\subsection{Explicit matrices}
To be able to simulate the amended protocol, we have to write down the matrix for the transformation $U$ explicitly. Any single-qubit gate $k \in \mathrm{SU(2)}$ can be written as
\begin{equation}
k(a_1,a_2,a_3) = \left( \begin{array}{cc}	e^{ia_1} \cos a_2	& e^{ia_3} \sin a_2 \\
															-e^{-ia_3} \sin a_2	& e^{-ia_1} \cos a_2
						\end{array} \right) \; .
\end{equation}
The explicit matrix for the non-local gate $A(c_1,c_2,c_3) = \exp [i (c_1\, \sigma_x \otimes \sigma_x + c_2\, \sigma_y \otimes \sigma_y + c_3\, \sigma_z \otimes \sigma_z)/2]$ is obtained by first finding the eigensystem of the hermitian operator $B := c_1\, \sigma_x \otimes \sigma_x + c_2\, \sigma_y \otimes \sigma_y + c_3\, \sigma_z \otimes \sigma_z$ and then applying
\begin{equation}
\label{eq:spectralconseq}
f(B) = \sum_j f(\lambda_j) |\lambda_j\rangle\langle\lambda_j| \; ,
\end{equation}
where $\lambda_j$ are the eigenvalues and $|\lambda_j\rangle$ the corresponding eigenvectors of the hermitian operator $B$. Equation (\ref{eq:spectralconseq}) is a direct consequence of the spectral decomposition theorem and holds for any analytic function $f$. In this case, $f(\cdot) = e^{i(\cdot)/2}$. The result is
\begin{multline}
A(c_1,c_2,c_3) = \\[3mm]
\begin{pmatrix}	e^{ic_3/2} \cos\left(\frac{c_1-c_2}{2}\right)	& 0	& 0	& i e^{ic_3/2} \sin\left(\frac{c_1-c_2}{2}\right) 			\\
											0	& e^{-ic_3/2} \cos\big(\frac{c_1+c_2}{2}\big)	& i e^{-ic_3/2} \sin\big(\frac{c_1+c_2}{2}\big)	& 0	\\
											0	& i e^{-ic_3/2} \sin\big(\frac{c_1+c_2}{2}\big)	& e^{-ic_3/2} \cos\big(\frac{c_1+c_2}{2}\big)	& 0	\\
											i e^{ic_3/2} \sin\big(\frac{c_1-c_2}{2}\big)	& 0	& 0	& e^{ic_3/2} \cos\big(\frac{c_1-c_2}{2}\big)
						\end{pmatrix} \; .
\end{multline}

\subsection{Simulation}
We simulate the progress of our protocol with different transformations $U$ and observe the QBER induced by Eve and Eve's mutual information on Alice's sifted key. Eve is assumed to employ the intercept-resend attack. She is allowed to choose between a projective measurement in either the $z$ or $x$ basis and not to perform any measurement individually for each of the two transmitted qubits. In the original BB84 protocol, it is clear that the $z$ and $x$ bases are the best measurement bases for Eve. In our augmented protocol, this is not necessarily true. For simplicity, however, we restrict Eve's measurements to these bases.

\subsubsection{Preliminary remarks}

Let $A$ denote the random variable that fully determines which pair of BB84 states $|a_1\rangle|a_2\rangle$ Alice constructs prior to the application of the non-local gate. That is, $A$ takes its values with uniform probability from the set $\{00,01,10,11,0+,0-,1+,\linebreak1-,+0,+1,-0,-1,++,+-,-+,--\}$. The physical state is obtained for each outcome $a_1a_2$ by surrounding the label with the bracket construct $|a_1a_2\rangle$. Let us re-label the outcomes with integers in the range $[0,15]$, in the order they are presented above, with the symbol $a$. For example, $a_1 = +$ and $a_2 = 1$ is expressed as $a=9$.

Let $E$ denote the random variable that gives the joint result of Eve's measurements, and let $e$ denote the outcome of $E$. The value of $e \in \{0,1,2,3\}$ is obtained by interpreting the separate results $e_1, e_2 \in \{0,1\}$ as a binary number $e_1e_2$ with $e_2$ as the least significant bit.

We may calculate Eve's mutual information on Alice's sifted key as her mutual information $I(A,E)$ on the random variable $A$. These two mutual informations are equal, which can be shown in exactly the same way as was done in Sec.~\ref{sec:bb84attacks} for one cbit and qubit in the original BB84 protocol. Due to the close similarities, the proof is not reproduced for the case of two cbits and qubits.
%Let $a$ denote the outcome of $A$, i.e., $a := a_1a_2$.

Because of finite computing resources, we use only the non-local gate $A(c_1,c_2,c_3)$ in our simulation. It is likely that using in addition the local gate $k_1$ benefits Alice and Bob, but it is also plausible that part of this gate commutes with the gate $A$ in the sense that Eve would be able to undo $k_1$ partially with single-qubit gates. This possibility is not investigated further in this Thesis.

\begin{figure}[hbt]
\vspace{5mm}
\begin{center}
\includegraphics[width=\textwidth]{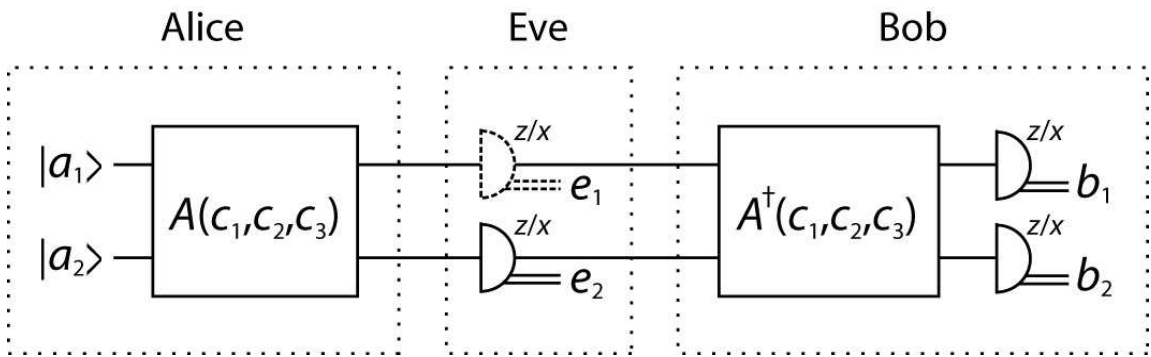}
\caption{The full quantum circuit of the proposed protocol and the attack for a qubit pair. The interleaving classical communication between Alice and Bob is not shown. The circuit is run a large number of times in each use of the protocol. The actions of each participant are enclosed in dotted boxes. A semicircle represents a projective measurement. Eve performs measurements but not necessarily on the first qubit. Symbols $e_1$ and $e_2$ denote Eve's measurement results, and $b_1$ and $b_2$ Bob's results which are assumed to contribute to the sifted key.}
\label{fig:protcirc}
\end{center}
\end{figure}

Figure \ref{fig:protcirc} shows the quantum transmission phase of the protocol and the attack we simulate as a quantum circuit for one qubit pair. In practice, Eve's measurement scheme may be such that the measured qubits are demolished. In this case, she creates new physical qubits in the logical state corresponding to her measurement result. This is equivalent to performing a non-demolishing projective measurement.

%tähän koko simulaation quantum circuit (even 1-qubittiportteineen?)

The protocol is sampled over a large number of different gates $A(c_1,c_2,c_3)$. We run a numerical \emph{Mathematica} code that records $I(A,E)$ and the QBER observed by Bob for Eve's allowed measurement bases and a given gate $A$. The algorithm is presented below. The code is run with different values of the parameters $c_1,c_2,c_3$ for the gate $A$. Each $c_i$ takes values in the interval $[0,\pi]$ with $\frac{\pi}{32}$ steps. That is, we sample the three-dimensional parameter space uniformly with $33^3 \approx 36000$ points. This is not an exhaustive survey of the possibilities of the use of a non-local gate, but as long as Eve obeys the presented assumptions, the obtained maximum of $I(A,E)$ holds for the given parameters $c_1,c_2,c_3$.

The following paragraphs describe phase-by-phase the algorithm used in the simulation of our protocol. The algorithm has not been optimized for performance, but instead kept in a form close to the underlying mathematics and physics. In the actual code, all calculation is done numerically. Every transmission state of the original BB84 protocol is assumed to occur with equal probability, and we only consider transmissions that contribute to the sifted key.

\subsubsection{Phase 1: Non-local transformation}
First, the gate $A(c_1,c_2,c_3)$ is applied to all 16 possible qubit-pair states used in the original BB84 protocol.
\begin{equation}
|\psi_{a_1a_2}\rangle = A(c_1,c_2,c_3) |a_1\rangle|a_2\rangle \; , \quad a_1,a_2 \in \{0,1,+,-\} \; .
\end{equation}
The qubit in the left slot, i.e., originally in state $|a_1\rangle$, is sent first.

\subsubsection{Phase 2: Eve's first measurement}
Eve may choose not to measure either of the two qubits. Since we apply a symmetric gate to the qubit pair, measuring only the first qubit is equivalent to measuring only the second qubit. Therefore, it suffices to simulate the protocol with Eve skipping the measurement only on the first qubit. If Eve does not measure either qubit, there is nothing to simulate.

If Eve has chosen to measure the first qubit, the measurement is calculated in the $z$ and $x$ bases. Based on Eq.~(\ref{eq:measprob}), the probability of measurement outcome $e_1$ is
\begin{equation}
\label{eq:eve1p}
p\left(E_1=e_1 | P_{\mathrm{E}1} = \epsilon_1, A=a\right) = \langle\psi_{a_1a_2}|\left(P^{\epsilon_1}_{e_1} \otimes I \right)|\psi_{a_1a_2}\rangle \; ,
\end{equation}
where $E_n$ is the random variable corresponding to the result of Eve's measurement of qubit $n \in \{1,2\}$ and $\epsilon_1$ is Eve's basis choice. The post-measurement state is
\begin{equation}
|\psi_{a_1a_2}(e_1,\epsilon_1)\rangle := \left(P^{\epsilon_1}_{e_1} \otimes I \right)|\psi_{a_1a_2}\rangle \big/ \sqrt{p(e_1 | \epsilon_1, a)} \; ,
\end{equation}
in accordance with Eq.~(\ref{eq:measstate}). However, if $p(e_1|\epsilon_1, a)=0$, we define
\begin{equation}
|\psi_{a_1a_2}(e_1,\epsilon_1)\rangle := \left( \begin{array}{cccc} 0 & 0 & 0 & 0 \end{array} \right)^{\mathrm{T}} \; .
\end{equation}
If Eve does not measure the first qubit, we define $|\psi_{a_1a_2}(e_1,\epsilon_1)\rangle := |\psi_{a_1a_2}\rangle$ and mark the probabilities of all outcomes as 1, which is mathematically inconsistent, but is later taken into account.

\subsubsection{Phase 3: Eve's second measurement}
Eve chooses a measurement basis for the second qubit and applies a projective measurement in the $z$ or $x$ basis. The probability of measurement outcome $e_2$ is
\begin{eqnarray}
\label{eq:eve2p}
&&p\left(E_2=e_2 | E_1 = e_1, P_{\mathrm{E}1}=\epsilon_1, P_{\mathrm{E}2}=\epsilon_2, A=a\right) \nonumber \\
&&= \langle\psi_{a_1a_2}(e_1,\epsilon_1) | \left(I \otimes P_{e_2}^{\epsilon_2} \right) |\psi_{a_1a_2}(e_1,\epsilon_1)\rangle \; ,
\end{eqnarray}
where $\epsilon_2$ is Eve's basis choice for the second measurement. If the result $e_1$ is impossible, the probability in Eq.~(\ref{eq:eve2p}) is correctly zero, since in this case the state vector is the zero vector. After both of Eve's measurements, Bob is in possession of the qubit pair which is in state
\begin{equation}
|\psi_{a_1a_2}(e_1,\epsilon_1,e_2,\epsilon_2)\rangle := \left(I \otimes P_{e_2}^{\epsilon_2} \right) |\psi_{a_1a_2}(e_1,\epsilon_1)\rangle \big/ \sqrt{p\left(e_2 | e_1, \epsilon_1, \epsilon_2, a\right)} \; .
\end{equation}
Again, if $p\left(e_2 | e_1, \epsilon_1, \epsilon_2, a\right) = 0$, we define the state to be the zero vector.

\subsubsection{Mutual information}

We allow Eve to choose the measurement basis independently for each qubit and calculate $I(A,E)$ for the different basis choices. Since the QBER also depends on the measurement basis, we must keep track of the results of all the choices to obtain a complete picture of Eve's capabilities. Because we consider the different basis choices separately, we may omit the explicit conditioning on the basis in all probabilities.

To be able to compare our results with other QKD protocols, we calculate Eve's information per bit. The mutual information of $A$ and $E$ given by Eq.~(\ref{eq:mi}) yields Eve's information on a two-bit entity. Thus the mutual information per bit is half of this, i.e.,
\begin{equation}
\label{eq:aeinfo2}
I(A,E) = \frac{1}{2} \left[ H(A) + H(E) - H(A,E) \right] \; .
\end{equation}
The entropy $H(A)$ is always
\begin{equation}
H(A) = - \sum_{j=0}^{15} p(a) \log p(a) = - \sum_{j=0}^{15} \frac{1}{16} \log \left(\frac{1}{16}\right) = 4 \; .
\end{equation}
According to Eq.~(\ref{eq:cond1}),
\begin{eqnarray}
&&p(e|a) \nonumber \\
&&= p(E=e_1e_2| A=a) \nonumber \\
&&= p(e_2 | e_1, \epsilon_1, \epsilon_2, a) p(e_1 | \epsilon_1, \epsilon_2, a) \; ,
\end{eqnarray}
the two factors of which have been calculated in Eqs.~(\ref{eq:eve1p}) and (\ref{eq:eve2p}). The entropy of Eve's variable is
\begin{eqnarray}
H_m(E)	& =	& - \sum_{e=0}^m p(e) \log p(e) \nonumber \\
		& \stackrel{(\ref{eq:totalprob})}{=}	& - \sum_{e=0}^m \left[\sum_{a = 0}^{15} p(e|a)p(a) \right] \log \left[\sum_{a = 0}^{15} p(e|a)p(a) \right] \nonumber \\
		& =	& - \frac{1}{16} \sum_{e=0}^m \left[\sum_{a = 0}^{15} p(e|a)\right] \left\{\log\left[\sum_{a = 0}^{15} p(e|a)\right] - 4 \right\} \; ,
\end{eqnarray}
where $m = 1$, if Eve measures only the second qubit, and $m=3$, if Eve measures both qubits. This is justified by noting that if the first measurement is not performed, all probabilities of $e_1$ are designated value 1 and $E_2$ does not depend on $E_1$, and thus $p(e|a) = p(e_2|a)$ given by Eq.~(\ref{eq:eve2p}).

The joint entropy of $A$ and $E$ is
\begin{eqnarray}
H_m(A,E)	& \stackrel{(\ref{eq:jointentr})}{=}		& -\sum_{e=0}^m \sum_{a = 0}^{15} p(e,a) \log p(e,a) \nonumber \\
				& \stackrel{(\ref{eq:condprob})}{=}	& -\sum_{e=0}^m \sum_{a = 0}^{15} p(e|a) p(a) \log \left[p(e|a) p(a)\right] \nonumber \\
				& =	& -\frac{1}{16} \sum_{e=0}^m \sum_{a = 0}^{15} p(e|a) \left[ \log p(e|a) - 4 \right] \nonumber \\
				& =	& \frac{1}{4} \sum_{e=0}^m \sum_{a = 0}^{15} p(e|a) - \frac{1}{16} \sum_{e=0}^m \sum_{a = 0}^{15} p(e|a) \log p(e|a) \; .
\end{eqnarray}
If $p(e|a)=0$ for some $e$ and $a$, we assign value zero to the term $p(e|a) \log p(e|a)$.

\subsubsection{Phase 4: Inverse non-local transformation}

Once Bob has received both qubits, he applies $A^{\dagger}(c_1,c_2,c_3)$ to the pair, and obtains the state
\begin{equation}
|\psi_{a_1a_2}^{\mathrm{Bob}}(e_1,\epsilon_1,e_2,\epsilon_2)\rangle := A^{\dagger}(c_1,c_2,c_3) |\psi_{a_1a_2}(e_1,\epsilon_1,e_2,\epsilon_2)\rangle \; .
\end{equation}
If Eve had not interfered with either qubit, it would be safe to write this as a product state.

\subsubsection{Phase 5: Bob's first measurement}
Bob projectively measures both qubits in correct bases. For the first qubit, the probability of result $b_1 \in \{0,1\}$ given $A = a_1a_2$ and $E = e_1e_2$ is
\begin{equation}
p(b_1|a_1a_2,e_1e_2) = \langle\psi_{a_1a_2}^{\mathrm{Bob}}(e_1,\epsilon_1,e_2,\epsilon_2)| \left(P_{b_1}^{\delta_{1,a}} \otimes I \right) |\psi_{a_1a_2}^{\mathrm{Bob}}(e_1,\epsilon_1,e_2,\epsilon_2)\rangle \; ,
\end{equation}
where 
\begin{equation}
\delta_{1,a} = \left\{ \begin{array}{ll}
	z & \textrm{if $a \in [0,7]$}\\
	x & \textrm{if $a \in [8,15]$\;.}
	\end{array} \right.
\end{equation}
If the result $e_1e_2$ is impossible, the probability is zero because the state vector is the zero vector. The probability of $b_1$ given only $A = a_1a_2$ is
\begin{equation}
p(b_1|a) \stackrel{(\ref{eq:probsum})}{=} \sum_{e=0}^m p(b_1,e|a) \stackrel{(\ref{eq:cond1})}{=} \sum_{e=0}^m p(b_1|a,e) p(e|a) \; .
\end{equation}
Bob's first measurement projects the qubit pair into state
\begin{equation}
|\psi_{a_1a_2}^{\mathrm{Bob}}(e_1,\epsilon_1,e_2,\epsilon_2,b_1)\rangle = P_{b_1}^{\delta_{1,a}}|\psi_{a_1a_2}^{\mathrm{Bob}}(e_1,\epsilon_1,e_2,\epsilon_2)\rangle \big/ \sqrt{p(b_1|a_1a_2,e_1e_2)} \; ,
\end{equation}
unless $p(b_1|a_1a_2,e_1e_2) = 0$, in which case the state is the zero vector.

\subsubsection{QBER of the first qubit}
If the transmission has no errors, the first measurement yields value 0 for $a \in \mathcal{A}_{10} = \{0,1,4,5,8,9,12,13\}$, and value 1 for $a \in \mathcal{A}_{11} = \{2,3,6,7,10,11,14,15\}$. Thus the QBER of the first qubit is, according to Eq.~(\ref{eq:QBER}),
\begin{equation}
\mathrm{QBER}_1 = \frac{1}{16} \left[ \sum_{a \in \mathcal{A}_{10}} p(B_1 = 1 |a) + \sum_{a \in \mathcal{A}_{11}} p(B_1 = 0 |a) \right] \; ,
\end{equation}
where $B_i$ is the random variable that yields the result of Bob's $i$th measurement.

\subsubsection{Phase 6: Bob's second measurement}
Bob measures the second qubit. The probability of getting result $b_2 \in \{0,1\}$ given $A = a_1a_2$ is
\begin{eqnarray}
p(b_2 | a)	& \stackrel{(\ref{eq:probsum})}{=}	& \sum_{b_1=0}^1 p(b_2, b_1 | a) \nonumber \\
				& \stackrel{(\ref{eq:cond1})}{=}		& \sum_{b_1=0}^1 p(b_2| b_1, a) p(b_1|a) \nonumber \\
				& \stackrel{(\ref{eq:probsum})}{=}	& \sum_{b_1=0}^1 \sum_{e=0}^m p(b_2, e| b_1, a) p(b_1| a) \nonumber \\
				& \stackrel{(\ref{eq:cond1})}{=}		& \sum_{b_1=0}^1 \sum_{e=0}^m p(b_2| b_1,e,a) p(e| b_1, a) p(b_1|a) \nonumber \\
				& \stackrel{(\ref{eq:cond4})}{=}		& \sum_{b_1=0}^1 \sum_{e=0}^m p(b_2| b_1,e,a) p(b_1| e, a) \frac{p(e|a)}{p(b_1|a)} p(b_1|a) \nonumber \\
				& =	& \sum_{b_1=0}^1 \sum_{e=0}^m p(b_2| b_1,e,a) p(b_1| e, a) p(e|a) \; .
\end{eqnarray}
The first factor in the term is obtained by calculating
\begin{equation}
p(b_2| b_1,e, a) = \langle\psi_{a_1a_2}^{\mathrm{Bob}}(e_1,\epsilon_1,e_2,\epsilon_2,b_1)| \left( I \otimes P_{b_2}^{\delta_{2,a}} \right) |\psi_{a_1a_2}^{\mathrm{Bob}}(e_1,\epsilon_1,e_2,\epsilon_2,b_1)\rangle \; ,
\end{equation}
where
\begin{equation}
\delta_{2,a} = \left\{ \begin{array}{ll}
	z & \textrm{if $a \in [0,3]$ or $a \in [8,11]$}\\
	x & \textrm{if $a \in [4,7]$ or $a \in [12,15]$\;.}
	\end{array} \right.
\end{equation}
Again, if result $b_1$ or result $e_1e_2$ is impossible, the probability is zero because the state is the zero vector.

\subsubsection{Total QBER}
After an error-free transmission, the second measurement yields value 0 for even values of $a$ and 1 for odd values of $a$. Therefore, the QBER of the second qubit is
\begin{equation}
\mathrm{QBER}_2 = \frac{1}{16} \left[ \sum_{a \mathrm{\ even}} p(B_2 = 1 |a) + \sum_{a \mathrm{\ odd}} p(B_2 = 0 |a) \right] \; .
\end{equation}
The total QBER is the average over the individual error rates:
\begin{equation}
\mathrm{QBER} = \frac{1}{2}\left(\mathrm{QBER}_1 + \mathrm{QBER}_2\right) \; .
\end{equation}

\subsection{Results}
We are interested in finding the maximum of Eve's mutual information on Alice's sifted key, $I(A,E)$, for a given QBER observed by Alice and Bob. In the analyzed attack, Eve has six different measurement configurations for each gate $A(c_1,c_2,c_3)$. She can measure both qubits in bases $zz$, $zx$, $xz$, or $xx$, or she can measure only the second qubit, in $z$ or $x$ basis. Due to the symmetry of the protocol with respect to the two transmitted qubits, the case where Eve measures only the first qubit needs no analysis---it is equivalent to measuring only the second one. Furthermore, using any of the six configurations, Eve can choose to interfere with only a fraction $0 \leq \xi \leq 1$ of the transmitted qubit pairs.

\begin{figure}[hbtp]
\vspace{5mm}
\begin{center}
\includegraphics[width=0.47\textwidth]{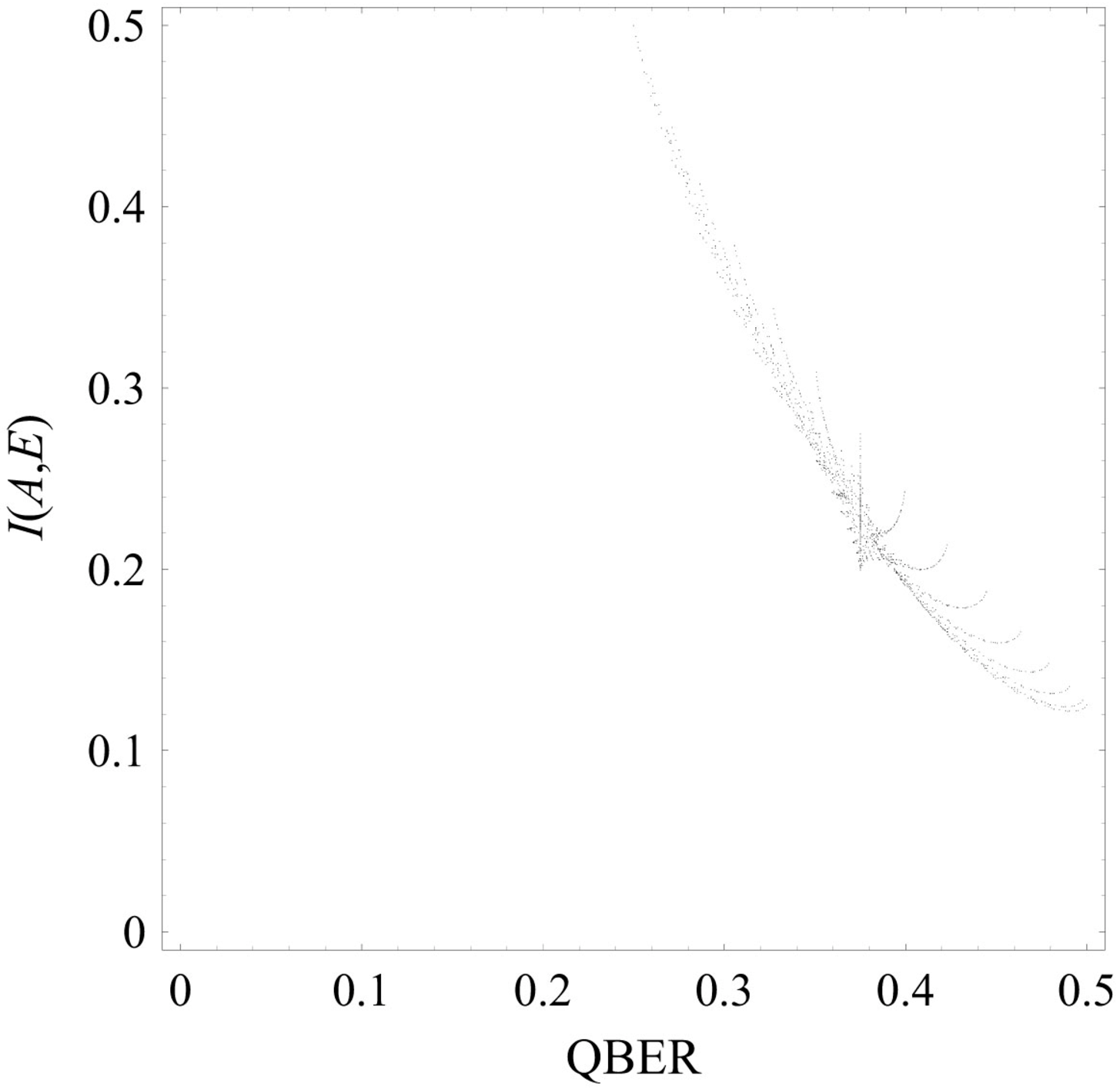}\quad \quad
\includegraphics[width=0.47\textwidth]{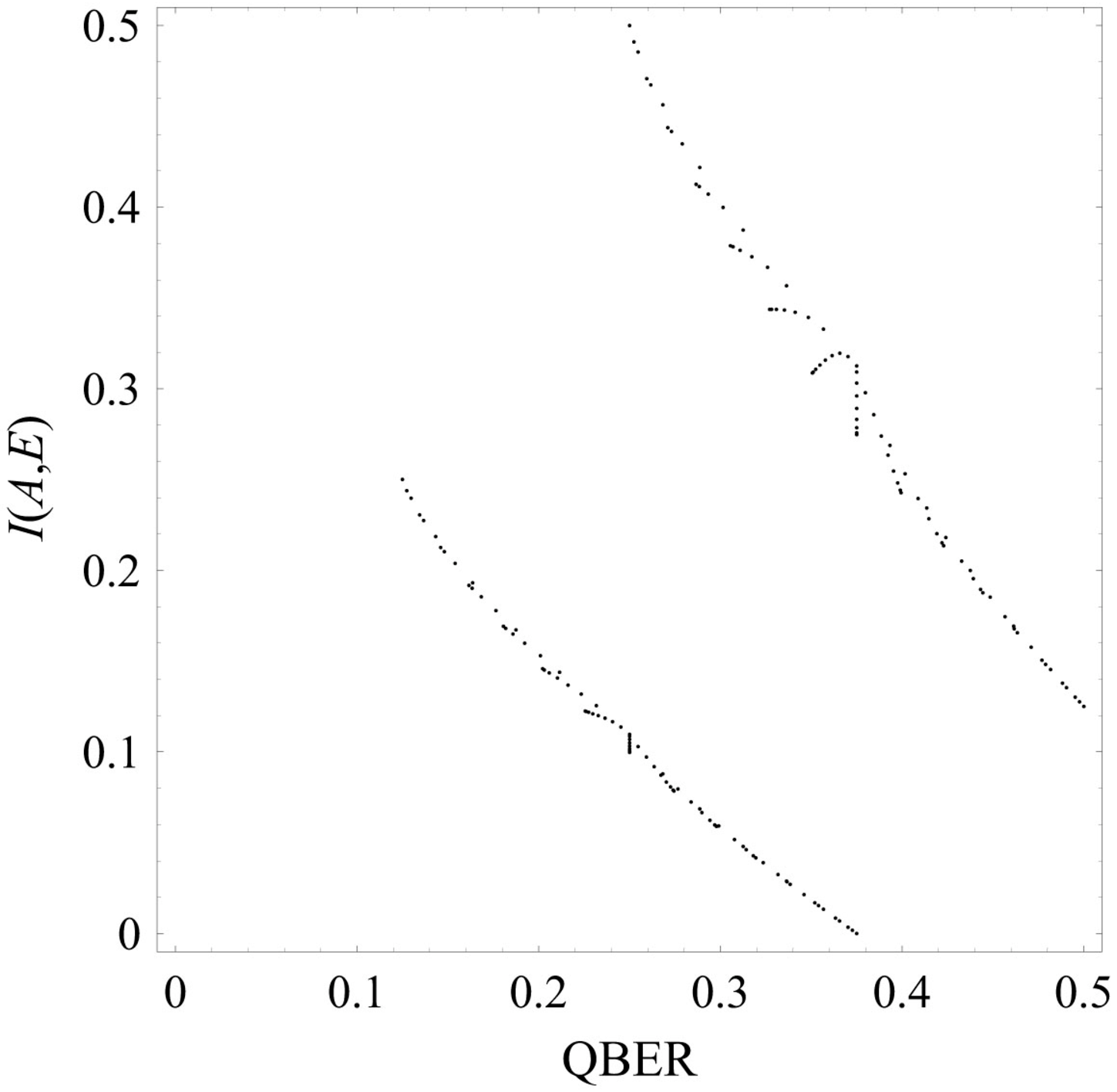}
\caption{Eve's mutual information per bit on Alice's sifted key and the corresponding QBER for the sampled values of $c_1,c_2,c_3$, and $\xi = 1$. The left panel shows the case where Eve measures both qubits and chooses $zx$ or $xz$ as her measurement bases. In the right panel, the lower set of points corresponds to Eve measuring only the second qubit in either basis, and the upper set to Eve measuring both qubits in the same basis. The upper envelope curve of the set in the left panel is the lower envelope curve for the upper set in the right panel.}
\label{fig:siivut}
\end{center}
\end{figure}

Figure \ref{fig:siivut} shows $I(A,E)$ as a function of QBER for the sampled parameter values in the different measurement configurations. The results are identical for bases $zx$ and $xz$, and very similar for bases $zz$ and $xx$, if Eve measures both qubits. If Eve measures only the second qubit, the results are very similar for both basis choices. The fraction $\xi = 1$ in all plots.

\begin{figure}[hbtp]
\vspace{5mm}
\begin{center}
\includegraphics[width=0.72\textwidth]{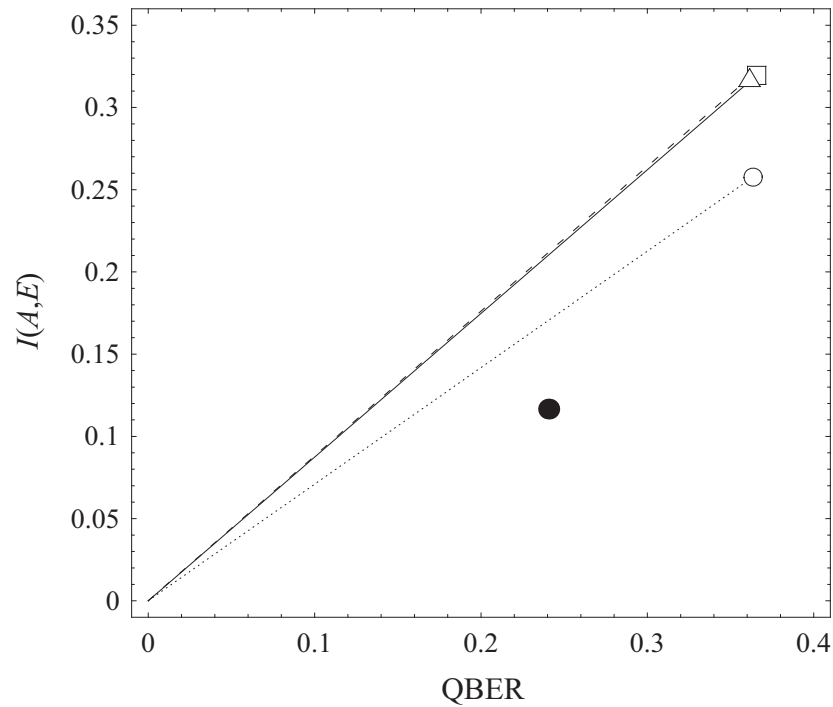}
\caption{Eve's information on Alice's key and the induced QBER for the gate $A\left(\frac{6 \pi}{32},\frac{25 \pi}{32},\frac{5 \pi}{32}\right)$. The filled circle corresponds to Eve measuring only the second qubit. The open symbols correspond to the cases where Eve measures both qubits: The circle denotes the $zx$ or $xz$ bases, the square bases $zz$, and the triangle bases $xx$. The mutual information and the QBER are slightly larger for the $zz$ than for the $xx$ choice. The solid, dashed, and dotted lines show Eve's information for $\xi \in [0,1]$ for bases $zz$, $xx$, and $zx$, respectively.}
\label{fig:7365}
\end{center}
\end{figure}

Which measurement configuration yields most information on Alice's key? For example, consider the situation for the gate $A\left(\frac{6 \pi}{32},\frac{25 \pi}{32},\frac{5 \pi}{32}\right)$, shown in Fig.~\ref{fig:7365}. The filled circle represents Eve measuring only the second qubit. This configuration provides least information and induces least errors. For the same QBER $\approx 0.24$, Eve obtains more information with any other configuration by adjusting $\xi$ properly, illustrated by the dashed and solid lines parametrized by $0 \leq \xi \leq 1$. Hence, measuring only one of the qubits does not provide maximal information. In fact, the same reasoning applies for any gate setting $c_1,c_2,c_3$, and we therefore ignore this measurement configuration in the following analysis.

\begin{figure}[hbtp]
\vspace{5mm}
\begin{center}
\includegraphics[width=0.85\textwidth]{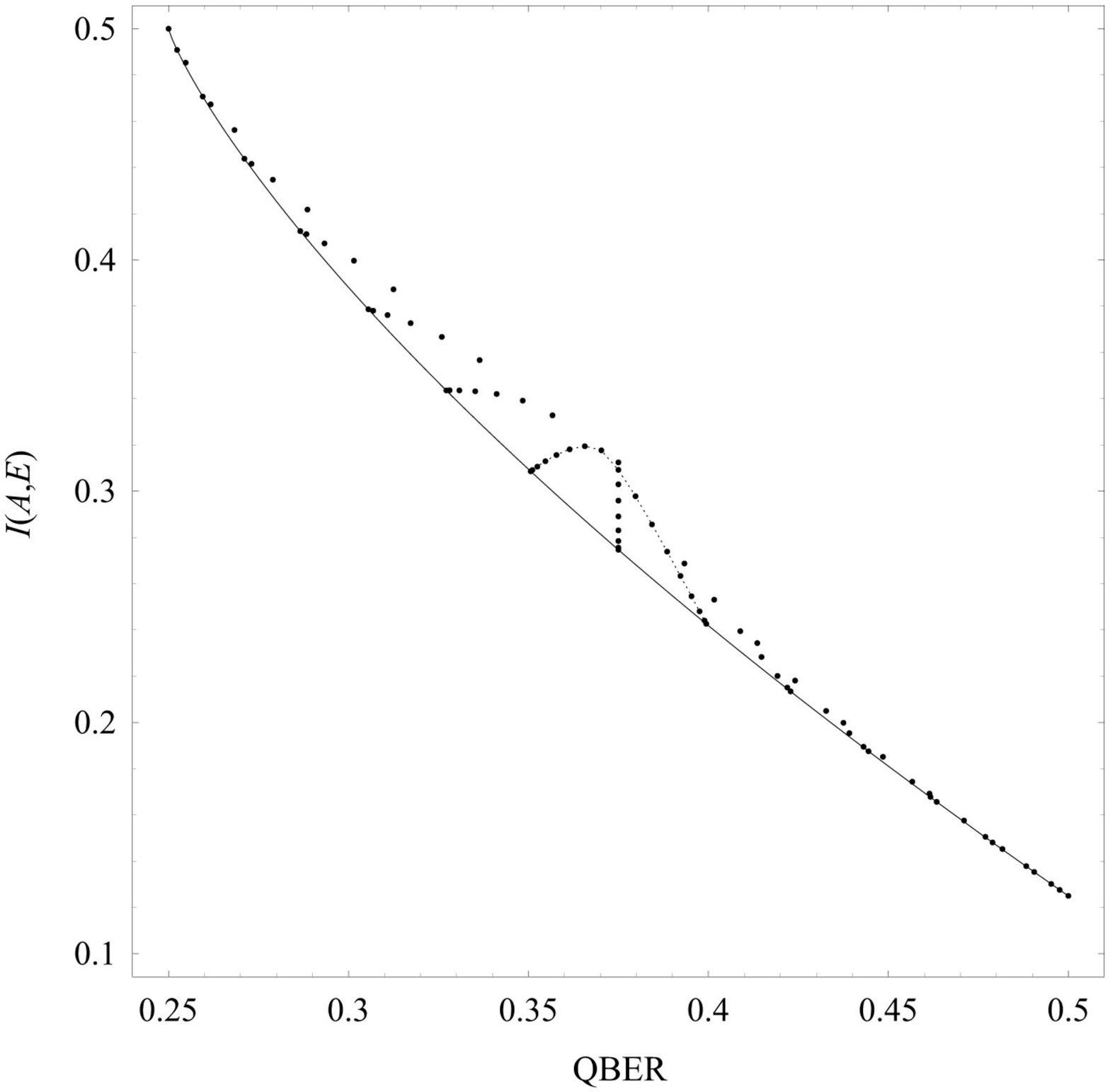}
\caption{Eve's information on Alice's key and the induced QBER for the sampled gates and $\xi=1$, given that Eve chooses the measurement configuration yielding the largest slope for the $\xi$-parametrized line. The solid line illustrates the lower envelope of the set, obtained by sweeping over $c_2$ for $c_1=c_3=0$. The dashed line is obtained by sweeping over $c_3$ for $c_1=0$ and $c_2=\frac{13 \pi}{16}$.}
\label{fig:miqsel}
\end{center}
\end{figure}

By measuring both qubits in the $z$ basis, denoted by the square in Fig.~\ref{fig:7365}, Eve maximizes her information as well as the QBER for the considered gate. Although measuring both qubits in the $x$ basis, denoted by the triangle, yields less information, the relative decrease in the QBER is larger. Thus, for any QBER up to that of the triangle, the $xx$ bases provide most information. That is, the slope of the $\xi$ line is larger for the $xx$ bases. Furthermore, as is shown in Fig.~\ref{fig:siivut}, for $\xi=1$, the induced QBER is always at least 25\%. Having observed a QBER this high, Alice and Bob would most likely abort the protocol. Hence, Eve should always adjust $\xi < 1$ such that the QBER is well below 25\%, and choose the configuration providing the largest slope for the line and thus maximal information. For any gate $A$, this configuration always involves measuring both qubits in either the $z$ basis or the $x$ basis, and hence we choose this to be Eve's configuration. This result is consistent with the symmetry of the gate $A$---there is no reason why it would be beneficial to employ different measurement bases for the qubits. Figure \ref{fig:miqsel} shows the mutual information as a function of QBER for the selected configuration and $\xi=1$. For each sampled gate, the maximal information is at most $0.011$ bits more than that given by the configuration corresponding to the largest slope.

The plot in Fig.~\ref{fig:miqsel} is generated as follows. For $\left(c_1,c_2,c_3\right) = (0,0,0)$ the mutual information is 0.5 and the QBER 25\%. This is consistent with the original BB84, since $A(0,0,0)$ is just the two-qubit identity transformation. The smallest information and the highest QBER, point $(0.5,0.125)$, is achieved with, e.g., the gate $A\left(0,\frac{\pi}{2},0\right)$. The lower envelope curve for the set of points in Fig.~\ref{fig:miqsel} is obtained by sweeping $c_2 \in \left[0,\frac{\pi}{2}\right]$ while keeping $c_1 = c_3 = 0$. Each of the concave arcs above the envelope are obtained by sweeping over $c_3$ for different values of $c_2$. For instance, if $c_1=0$ and $c_2 = \frac{13 \pi}{16}$, increasing $c_3$ from 0 to $\frac{\pi}{2}$ produces the dashed arc in Fig.~\ref{fig:miqsel}. For different values of $c_1$, the sweeps over $c_2$ and $c_3$ yield arcs of different shape in the same set of points. We thus observe that adjusting $c_1$ is redundant, the same effect is achieved by an appropriate choice of $c_2$ and $c_3$.

\begin{figure}[hbt]
\vspace{5mm}
\begin{center}
\includegraphics[width=0.6\textwidth]{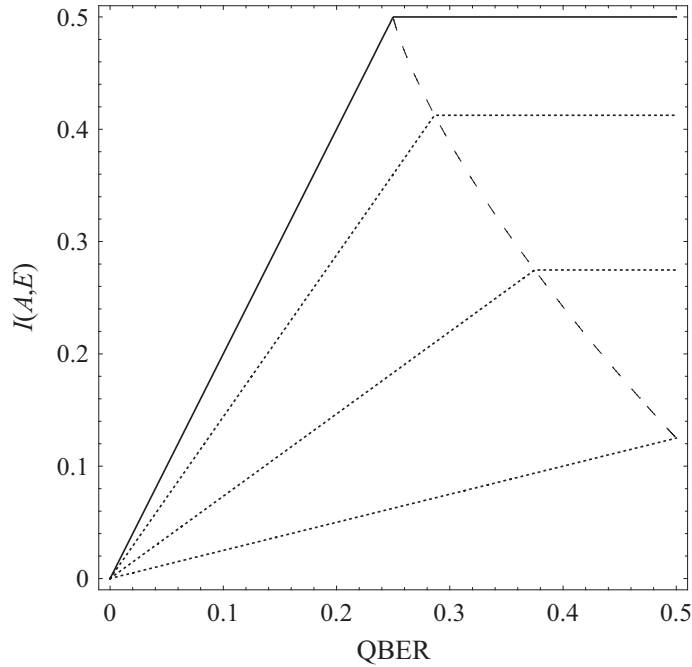}
\caption{Eve's maximal mutual information and the induced QBER for BB84 (solid line) and for our protocol with gate settings $c_2 = \frac{\pi}{8}, \frac{\pi}{4}, \frac{\pi}{2}$ and $c_1 = c_3 = 0$, denoted by the uppermost, middle, and lowest dashed line, respectively. Eve employs an IR attack. The sparsely dashed line denotes the envelope curve of the set shown in Fig.~\ref{fig:miqsel}.}
\label{fig:vertailu}
\end{center}
\end{figure}

Let us compare our protocol with the original BB84. Figure \ref{fig:vertailu} shows Eve's maximal information as a function of QBER for BB84 and for three representative gates in our protocol. The oblique lines are obtained by varying $\xi \in [0,1]$. The solid line denotes BB84 and the densely dashed lines our protocol for settings $c_1=c_3=0$ and $c_2 = \frac{\pi}{8}, \frac{\pi}{4}, \frac{\pi}{2}$. Our protocol provides Eve less information and induces more errors in the sifted key for a given $\xi$. For instance, by employing gate $A(0,\frac{\pi}{8},0)$, Eve's information is decreased by $0.0875$ bits and the QBER increased by only 0.037, for $\xi=1$. In BB84, the incoherent attack provides approximately 0.13 bits more information than the IR attack for a 20\% QBER, as is shown in Fig.~\ref{fig:bb84miqber}. In our protocol, Eve's information for the same QBER can be reduced from 0.4 to 0.05---much more than what is gained by applying an incoherent attack in BB84. The difference is significant for any QBER less than 25\%.

The error correction phase provides Eve further information. Equation (\ref{eq:errcorrbits}) gives a lower bound on the number of bits Alice and Bob need to exchange to correct the errors, with an error probability of $p$ = QBER in each bit. The bound is valid for an error process affecting each bit individually, but in our protocol, the errors in the bits obtained from one qubit pair are correlated due to entanglement. However, we can apply the bound to our protocol as well, since correcting pairwise correlated errors cannot be more demanding than correcting independent errors. That is, Alice and Bob can treat the errors as uncorrelated. Because the number of exchanged bits only depends on the QBER, the differences of the lines in Fig.~\ref{fig:vertailu} remain the same after adding to them the information provided by the error-correction step. Thus, for a given QBER, our protocol provides Eve strictly less information with $c_2$ and $c_3$ chosen properly, assuming that Eve uses the described IR attack.
%The graph on the right in Fig.~\ref{fig:vertailu} shows Eve's combined information from both sources for BB84 and for our protocol with the same gates as on the left. In BB84, Eve achieves full information on the key at $\mathrm{QBER} \approx 17\%$.

%muutama ote ja sit error correctionilla -> turhaa
%ec kaava pätee iid erroreille, mutta voidaan käyttää ylärajana/arviona

Which gate $A\left(c_1,c_2,c_3\right)$ should Alice and Bob choose? The answer is complicated by the fact that any practical implementation of the protocol includes a quantum channel with a finite error rate. Eve's interference acts as an approximate, although poor, model for the noise in the quantum channel. While the choice $A\left(0,\frac{\pi}{2},0\right)$ limits Eve's information most effectively for any QBER, it also presumably amplifies the noise in the quantum channel the most. Whether this amplification is tolerable depends on the noise. However, since Eve's information decreases rapidly compared to the increase in QBER in the upper part of the envelope curve shown in Fig.~\ref{fig:vertailu}, it is plausible that even for a very noisy channel a non-identity gate benefits Alice and Bob. Moreover, the noise of a practical quantum channel decreases as technology advances.

For instance, assume that Alice and Bob employ the gate $A\left(0,\frac{\pi}{2},0\right)$ corresponding to the lowest dashed line in Fig.~\ref{fig:vertailu}. Assume also that the inherent noise of the used quantum channel results in a QBER of at most 10\%. If Eve's interference is used as a model for the noise, the QBER is doubled by the gate, since QBER = $\xi/4$ without the gate, and QBER = $\xi/2$ with the gate. Hence, we assume that if the gate is employed, Alice and Bob must accept a 20\% QBER. Without the gate, Eve's maximal information can be limited to 0.2 bits, and with the gate to 0.05 bits. That is, the gate reduces Eve's information to 1/4 of that in BB84 even if the noise is taken into account. Exactly which gate to employ depends on the actual noise of the realized quantum channel, however.

%oletetaan että eve valitsee aina jyrkimmän slopen
%pistejoukko = set of points

%\small{Miten laskettiin ja mita kun Evella kaytossa Z- ja X-kannat intercept-resendissa. \\Mathematica-koodi esitettyna pseudokoodina? Liiteeksi?}
%\small{Tulokset edellisista}

\section{EPR-pair attack}
Since Alice and Bob utilize entanglement to keep their key a secret, it is an intuitive idea for Eve to also make use of this resource. One way of taking advantage of entanglement in eavesdropping is to send qubits of an EPR pair, defined in Eq.~(\ref{eq:eprcharlie}), to Bob. Let us label the qubits of the EPR pair as
\begin{equation}
\label{eq:eprpairlab}
|\Phi^+\rangle = \frac{1}{\sqrt{2}} \left(|0\rangle_1|0\rangle_2 + |1\rangle_1|1\rangle_2\right) \; .
\end{equation}
 After intercepting the first qubit of the transmitted pair, Eve sends qubit 1 of an EPR pair to Bob while keeping qubit 2 for herself. Bob acknowledges the reception of the first qubit, and Alice sends the second qubit which Eve also intercepts. Eve can, e.g., measure the intercepted qubits, and based on the result, apply a single-qubit transformation to the second qubit of the EPR pair which she then sends to Bob. More complicated transformations involving the intercepted qubits are also possible. Eve has thus sent Bob a qubit pair in an entangled state one qubit at a time, and has partial control over the state of the pair after learning the measurement results for both intercepted transmissions.

For instance, assume that Alice and Bob have chosen $U_1 := \textrm{CNOT}\left(H \otimes I\right)$ as their two-qubit unitary transformation. CNOT is the non-local controlled-not operation which transforms input states as
\begin{equation}
\textrm{CNOT:\;} \left\{
\begin{array}{lcl}
|00\rangle	& \to	& |00\rangle \\
|01\rangle	& \to	& |01\rangle \\
|10\rangle	& \to	& |11\rangle \\
|11\rangle	& \to	& |10\rangle
\end{array}
\nonumber
\right.
\end{equation}
and $H$ is the Hadamard gate that transforms $|0\rangle \to (|0\rangle + |1\rangle)/\sqrt{2}$ and $|1\rangle \to (|0\rangle - |1\rangle)/\sqrt{2}$. The only BB84 states $|a_1\rangle|a_2\rangle$ not resulting in a tensor product state with the application of $U_1$ are states $|00\rangle, |01\rangle, |10\rangle, |11\rangle$. These states are transformed into the Bell states by $U_1$ as
\begin{eqnarray}
U_1|00\rangle	& =	& |\Phi^+\rangle = (|00\rangle + |11\rangle)/ \sqrt{2} \; , \\
U_1|01\rangle	& =	& |\Psi^+\rangle = (|01\rangle + |10\rangle)/ \sqrt{2} \; , \\
U_1|10\rangle	& =	& |\Phi^-\rangle = (|00\rangle - |11\rangle)/ \sqrt{2} \; , \\
U_1|11\rangle	& =	& |\Psi^-\rangle = (|01\rangle - |10\rangle)/ \sqrt{2} \; .
\end{eqnarray}
%In this case, on average 1/4 of the qubit pairs travel from Alice to Bob in a maximally entangled Bell state \cite{gisin}, and the rest in unentangled states.

Assume that Eve has some way of knowing if Alice uses the $z$ basis for the initial state $|a_1\rangle|a_2\rangle$. This capability would, of course, severely compromise the security of any BB84-based QKD protocol. Nevertheless, let us demonstrate how Eve could in this case eavesdrop on the transmission and still preserve its entanglement. If Alice has chosen some other combination of bases than $zz$ for the qubits, Eve attacks the unentangled transmission using any strategy suitable for the original BB84 protocol. However, if Alice's choice of bases is $zz$, Eve knows to expect a transmission of a qubit pair in one of the Bell states, and does the following.

%Note that Bob has not yet measured the EPR-pair qubit in his possession because he is waiting for the second qubit of the transmitted pair.
Eve intercepts and stores the first transmitted qubit in short-term quantum memory, and immediately sends qubit 1 of the pair in Eq.~(\ref{eq:eprpairlab}) to Bob. She then intercepts and stores the second transmitted qubit and is thus in possession of the transmitted qubit pair as well as the second qubit of the EPR pair. Eve undoes $U_1$ by applying $U_1^{\dagger} = (H \otimes I) \mathrm{CNOT}$ to the intercepted pair, after which she measures the qubits in $z$ basis, thus recovering $a_1$ and $a_2$ exactly. Based on the result, Eve chooses a single-qubit gate
\begin{equation}
E_{a_1a_2} = I \left(\sigma_x\right)^{a_2} \left(\sigma_z\right)^{a_1} \; ,
\end{equation}
which she applies to qubit 2 of the EPR pair. This transforms the pair to the state that Alice transmitted, i.e., $\left(I \otimes E_{a_1a_2} \right)|\Phi^+\rangle = U_1 |a_1\rangle|a_2\rangle$. Eve sends the second qubit to Bob, who applies $U_1^{\dagger}$ to the qubit pair. Bob measures the qubits in the $z$ basis, and recovers $a_1$ and $a_2$. That is, Bob observes a zero QBER while Eve has full knowledge about the key bits Alice and Bob established, given that Alice used the $zz$ basis. Note that this result only applies if Alice's basis choices are available to Eve at the time of the quantum transmission---a feature that would render the original BB84 protocol useless. The utilization of entanglement in eavesdropping is not discussed further in this Thesis.

%\section{Applicability}

%---
%\small{Jos ehditaan saada, niin Even incoherent attack tata protokolla vastaan.} -> ei ehditä

%% file: conclusions.tex
\chapter{Conclusions}
\label{ch:conclusions}
%vai Discussion

%mitä tehtiin, miten tehtiin, mikä oli tulos, mitä kannattaisi tehdä seuraavaksi

We have introduced and studied the properties of a novel BB84-based protocol for quantum key distribution. The proposed protocol utilizes entanglement of the transmitted quantum states to provide advantage against an eavesdropper. We derived a practical model for the entangling transformation, and simulated the protocol numerically over a subset of the transformations. We considered the security of the protocol under an intercept-resend attack. In the simulation, we recorded the information of the eavesdropper on the established key and the quantum bit error rate induced by the attack.

We find that entangling the states of the transmitted qubit pairs properly yields significant advantage to the legitimate users. The maximal mutual information available to an eavesdropper can be controlled in the range from 0.125 to 0.5 bits. Decreasing the maximal information increases the quantum bit error rate to at most 50\%. For a given quantum bit error rate below 25\%, the maximum information that an intercept-resend attack provides is reduced by a factor of eight. Since eavesdropping causes more disturbance to the quantum transmission than in the original protocol, an eavesdropper is detected more easily. In other words, for a given error rate, an eavesdropper must reduce his or her interference with the quantum channel, and thus acquire less information.

In practice, the entangling transformation also amplifies the inherent noise in the realized quantum channel. The transformation acts as controllable leverage in the protocol---while it limits the information of an eavesdropper, it also amplifies the effects of noise and eavesdropping. Thus, it may not be practical to reduce the maximal information of an eavesdropper to its minimum, since the noise may be amplified too much. However, the amplification of the noise does not reduce the benefits of the protocol substantially. Furthermore, as technology advances, the inherent noise of a quantum channel can be decreased. Studying the amplification of a channel noise in detail is a possible topic for future research.

In our protocol, an eavesdropper is assured to have only one-by-one access to the transmitted particles. We studied how much this, in general, restricts the capability of the eavesdropper of obtaining the transmitted entangled state. We maximized numerically the minimal error in approximating an entangled two-qubit state with a product state. We found that the maximum of the minimal error in the approximation is 0.673. Thus, even if the eavesdropper was equipped with a perfect quantum cloner, the proposed protocol would impose a significant hindrance to him or her.

In addition, we described a novel attack type against the proposed protocol and showed that it enables an eavesdropper to imitate the entanglement in the transmission. In this so-called EPR-pair attack, the eavesdropper captures the transmitted states and replaces them with the halves of an EPR pair. We showed that in a special case and under strong assumptions on the capabilities of the attacker and the entangling transformation, this attack allows the eavesdropper to gain full knowledge on the generated key. Note that these assumptions would render BB84 useless. Detailed investigation of this attack type is left for future research.

Suggestions for topics of future research also include the following. The security of the protocol could be analyzed in the presence of an attacker with more capabilities than what we have allowed. Firstly, the attacker can be allowed to measure the transmissions individually in any basis, implementable by allowing the use of arbitrary single-qubit gates, and to adjust each measurement based on previous results. Secondly, the security could be analyzed under an optimal incoherent attack, and, if possible, in the case of a collective attack. If the advantage is not lost under the most general attack, entangling the states of more than two particles would probably create an even larger advantage, since the dimension of an $N$-qubit state increases exponentially in $N$.

%\small{\begin{verbatim}future research: looking into discarded qubits,\end{verbatim}  \begin{verbatim}separating qubit #1 and #2 in qber and in info, using d>2 qubit blocks, application\end{verbatim}  \begin{verbatim} to 2- and 6-state protocols?\end{verbatim}}
%lisää analyysiä EPR-pari-hyökkäyksen potentiaalista
%evelle yksiqubittigatet
%tuleeko liika noisea gaten käytöstä?
%vipuvaikutus: herkistää protokollan myös viattomille virheille

%\small{Ja mika nyt sitten lopulta olikaan protokollan hyoty. (Hyotya perus-BB84:aan tiedetaan olevan kun Eve IR ja Z-X-kannassa, paihittaako tama protokolla muita BB84-virityksia.) Eve ainakin havaitaan helpommin koska yhdenkin qubitin sorkkiminen hairitsee molempia.}

%Because $U_d$ is unitary, its inverse can be calculated fast. itsestäänselvää

%entanglement measure for transmitted groups

%alicen ja bobin kandee valita kannat samoiksi qubittiparin qubiteille, muuten osuu yksiin vain alle puolessa tapauksista (tod.näk. = (1/2)*(1/2) = 1/4) -> EIJEI: U tehdään kannanvalintojen välissä => voi hyvin olla että toinen mätsää ja toinen ei, se on ihan ok.
%entäs kun eve hyökkää sellaisen parin kimppuun josta vain toinen qubitti päätyy sifted keyhin? -> eve saa tietää kumpi qubitti diskataan. symmetrian takia info pysyy samana. analyysi pätee.
%The protocol requires the use of high-fidelity two-qubit gates and short-term quantum memory, in addition to the requirements of the BB84 protocol.